\documentclass{aa} 

\usepackage{graphicx}
\usepackage{color}
\usepackage{float}

\begin{document}

\title{The chemical evolution of  self-gravitating primordial disks}
\titlerunning{The chemical evolution of  self-gravitating primordial disks} 
\authorrunning{Schleicher et al.}

\author
  {Dominik R.\,G. Schleicher
  \inst{1}
  \and
  Stefano Bovino
  \inst{2}
  \and
  Muhammad A. Latif
  \inst{3,4}
  \and
  Andrea Ferrara
  \inst{5,6}
  \and
  Tommaso Grassi
  \inst{7,8}
  }

\institute{Departamento de Astronom\'ia, Facultad Ciencias F\'isicas y Matem\'aticas, Universidad de Concepci\'on, Av. Esteban Iturra s/n Barrio Universitario, Casilla 160-C, Concepci\'on, Chile;
\email{dschleicher@astro-udec.cl}
\and Hamburger Sternwarte, University of Hamburg, Gojenbergsweg 112, 21029 Hamburg, Germany
\and
Sorbonne Universit\'es, UPMC Univ Paris 06, UMR 7095, Institut d'Astrophysique de Paris, F-75014, Paris, France
\and
CNRS, UMR 7095, Institut d'Astrophysique de Paris, F-75014, Paris, France
\and 
Scuola Normale Superiore, Piazza dei Cavalieri 7, 56126 Pisa, Italy
\and
Kavli Institute for the Physics and Mathematics of the Universe (WPI), Todai Institutes for Advanced Study, the University of Tokyo, 5-1-5 Kashiwanoha, Kashiwa, 277-8583, Japan
\and
Centre for Star and Planet Formation, Natural History Museum of Denmark, \O ster Voldgade 5-7, DK-1350 Copenhagen, Denmark
\and
Niels Bohr Institute, University of Copenhagen, Juliane Maries Vej 30, DK-2100 Copenhagen, Denmark
}

\date{\today}


\abstract{Numerical simulations  show the formation of self-gravitating primordial disks during the assembly of the first structures in the Universe, in particular during the formation of Pop.~III and  supermassive stars. Their subsequent evolution is expected to be crucial to determine the mass scale of the first cosmological objects, which depends on the temperature of the gas and the dominant cooling mechanism. Here, we derive a one-zone framework to explore the chemical evolution of such disks and show that viscous heating leads to the collisional dissociation of an initially molecular gas. The effect is relevant on scales of $10$~AU ($1000$~AU) for a central mass of $10$~M$_\odot$ ($10^4$~M$_\odot$) at an accretion rate of $10^{-1}$~M$_\odot$~yr$^{-1}$, and provides a substantial heat input to stabilize the disk. If the gas is initially atomic, it remains atomic during the further evolution, and the effect of viscous heating is less significant. The additional thermal support is particularly relevant for the formation of very massive objects, such as the progenitors of the first supermassive black holes. The stabilizing impact of viscous heating thus alleviates the need for a strong radiation background as a means of keeping the gas atomic.}

\maketitle

\section{Introduction}
The first stars in the Universe are expected to form from a primordial gas in minihalos of $10^5-10^6$~M$_\odot$ at $z=20-30$ \citep{Abel02, Bromm03, Yoshida08}. Their formation process is often explored from cosmological initial conditions, following the gravitational collapse over many orders of magnitude, to scales of astronomical units (AU) or even below. Due to the characteristic temperatures of $\sim300$~K in the primordial gas, the first stars are expected to be considerably more massive than present day stars. Adopting a temperature of $\sim10$~K in present-day molecular clouds, one may expect an increase in mass by a factor of $30^{1.5}\sim140$ based on simple Jeans arguments.

The actual masses and  their distribution depend however on the details of the star formation process, which are only partly understood. A central question concerns the role of fragmentation, which may lead to the formation of a stellar cluster rather than a central star \citep{Clark08}. The latter requires the formation of a self-gravitating disk, and its subsequent fragmentation via gravitational instabilities. Indeed, simulations starting from cosmological initial conditions have now confirmed the formation of self-gravitating disks in minihaloes, as well as their fragmentation \citep{Clark11, Greif11, Greif12, Latif13PopIII, Bovino13c, Susa14}. While these simulations followed the evolution for up to $\sim1000$~years depending on resolution, the Kelvin-Helmholtz timescale for a protostar to reach the main sequence is  of the order of $\sim10^6$~years, implying that no final conclusions can be drawn on the properties of the stellar system, and that the simulations so far have only explored the evolution of the disks at very early stages. An exception in this respect are the studies by \citet{Hosokawa11} and \citet{Hirano14}, who have followed the formation of primordial stars until reaching the main sequence, however at the price of a 2D-approximation. The latter  provides a hint towards the final mass distribution, in particular if fragmentation is not efficient or if the resulting clumps efficiently merge with the central object. For a further discussion concerning the state-of-the-art of current simulations exploring Pop.~III star formation, we refer to the recent review by  \citet{Greif15}.

In  massive primordial halos of $10^7-10^8$~M$_\odot$, the formation of very massive objects with $\sim10^5$~M$_\odot$ has been suggested to occur \citep[e.g.][]{Bromm03, Koushiappas04, Lodato06, Begelman09, Natarajan11b, Volonteri12}, as potential progenitors of the observed supermassive black holes at $z\sim6-7$ \citep{Fan01, Fan04, Fan06, Mortlock11, Banados14}. In the presence of a strong photo-dissociation background, particularly high temperatures of the gas can be achieved \citep[e.g.][]{Omukai01, Schleicher10b}, therefore favoring the formation of very massive primordial objects. Simulations have shown that self-gravitating disks form in haloes cooling via atomic hydrogen \citep{Regan09, LatifBH, Latif13BHmass, Prieto13, Regan14, Becerra15}, where central masses of $10^5$~M$_\odot$ can be reached within $10^4$~years after the initial collapse \citep{LatifBH}.  

The black hole mass function resulting from such a direct collapse has been derived by \citet{Ferrara14} using a semi-analytical approach, considering stellar evolution models for rapidly accreting protostars \citep{Hosokawa13, Schleicher13b}. Even for more moderate radiation fluxes, massive stars with $10^3-10^4$~M$_\odot$ may still form \citep{Latif14Star}, and the rotational support in disks was found to increase for an increasing amount of molecular cooling \citep{LatifV15}. Also in this context, it is therefore important to understand the long-term evolution of self-gravitating disks both in the atomic and molecular cooling regime. 

To complement such numerical investigations, analytical studies have been pursued to investigate the properties of primordial self-gravitating disks. For instance, \citet{Lodato06} have discussed the role of self-gravitational instabilities and fragmentation during the formation of a massive central black hole, which may be accompanied with a starburst in the self-gravitating disk. \citet{Inayoshi14} have explored the fragmentation properties of a self-gravitating disk cooling via atomic hydrogen, concluding that fragments forming via the gravitational instabilities would subsequently merge with the central object. \citet{Latif15Disk} have extended the model into the molecular cooling regime, finding that the merging of clumps is likely still efficient, as the characteristic migration timescales of the clumps are shorter than the timescales of protostellar contraction. In the regime of very massive stars or high accretion rates, the resulting viscous heating may even lead to a transition from the molecular to the atomic cooling regime \citep{Latif15}. 

\citet{Ferrara13} have explored the stability of self-gravitating protogalactic disks in the presence of metals, finding that these are subject to strong fragmentation. The latter may provide an obstacle for the formation of massive black holes, as suggested by \citet{Mayer10, Mayer14}. In the context of self-gravitating protostellar disks, the role of metal and dust cooling has been explored by \citet{Tanaka14}, suggesting that they may considerably enhance fragmentation. Similarly, the results of 3D simulations show that fragmentation is triggered by the additional cooling, especially for simulations that are evolved beyond the formation of the first peak. Such behavior has been observed for instance by \citet{Clark08, Shrader14, Bovino14, Peters14} and \citet{Shrader15}. 

A substantial debate at this point concerns the question of whether a purely primordial and atomic gas is required for the formation of supermassive black holes, as often assumed in statistical predictions for the high-redshift black hole population \cite[e.g.][]{Dijkstra08, Dijkstra14, Agarwal14}. The typical mechanism considered here is the photo-dissociation of molecular hydrogen by Lyman-Werner radiation, as explored e.g. by \citet{Shang10}, which is typically parametrized through a critical radiation flux $J_{\rm crit}$. Over the last years, it has been shown that the critical flux can be considerably enhanced by considering realistic stellar spectra \citep{Sugimura14, Agarwal15}, as well as by the dynamics in 3D simulations, where the initial ionization degree is enhanced by shocks and therefore requires an even stronger radiation background to prevent the formation of molecular hydrogen \citep{Latif14UV, Latif15X}. The simulations suggest a value of  $J_{\rm crit}$ of up to $10^5$. For the production of the observed population of $z\sim6$ black holes, this value is considerably too high, even when considering chemical uncertainties of at most a factor of $5$ \citep{Glover15}.

However, it is by no means clear whether such a critical radiation flux is in fact required for the formation of massive objects. For instance, \citet{Begelman09} suggested that massive objects could also form from a molecular gas in the presence of self-gravitating instabilities. Simulations by \citet{Latif14Star} confirmed that massive objects of $\sim10^3-10^4$~M$_\odot$ still form for radiation fluxes below $J_{\rm crit}$, and accretion rates of $\sim10^{-1}$~M$_\odot$~yr$^{-1}$ can be maintained even for a moderate radiation background \citep{LatifV15}. In addition, it is conceivable that an atomic gas forms through alternative pathways. For instance, \citet{Inayoshi12} considered the dissociation of the molecular gas by shocks, and showed that the gas would then remain atomic during the further collapse, while \citet{Sethi10} and \citet{Borm13} considered a similar effect due to the dissipation of magnetic energy. 

More recently, we  found indications that viscous heating in the presence of  rotation can substantially heat up the interior of a self-gravitating disk, leading to a transition towards an atomic cooling regime \citep{Latif15}. {The latter can be understood as the viscous heating rate, which strongly depends on the angular velocity of the gas, considerably increases towards smaller scales and will ultimately exceed the cooling rate of a molecular gas.} While previous investigations were predominantly exploring a free-fall collapse using one-zone models  \citep[e.g.][]{Omukai01, Omukai05, Glover08, Glover15} or the modeling of the early stages of disk formation \citep[e.g.][]{Regan09,Clark11, Greif11, Greif12, Latif13PopIII, LatifBH, Bovino13c, Prieto13, Regan14, Susa14, Becerra15}, we explore here the chemical and thermal evolution of such disks at their later stages, in particular during the presence of a central massive object. For this purpose, we will consider different disk models and different chemical initial conditions, representing both an initially atomic and an initially molecular gas. For the modeling of the chemistry, we employ the publicly available chemistry package \textsc{krome}\footnote{Webpage \textsc{krome}: http://kromepackage.org/} developed by \citet{Grassi14}.

The outline of this paper is  as follows. In section~\ref{theory}, we describe the one-zone model for the evolution in a self-gravitating disk in a Lagragian frame  as well as the model for the chemical evolution. In section~\ref{molecular}, we present our results for the evolution of disks with an initially molecular gas, and section~\ref{atomic} contains the results for initially atomic disks. A summary and discussion are presented in section~\ref{summary}.

\section{Theoretical framework}\label{theory}
In the following, we outline a one-zone model to describe the chemical evolution in the mid plane of primordial disks. For this purpose, we assume that the disk is axisymmetric and stationary in an Eulerian reference frame. We then consider a ring at radius $R$ with mass $dM$ and width $dR$, implying a surface density $\Sigma=dM/(2\pi R dR)$. In the following, we will consider how the gas in the ring moves inward by adopting a Lagrangian frame of reference and  follow its evolution. While the disk is stationary in the Eulerian frame, the gas ring will evolve when moving inward, implying that the density, temperature and chemical abundances will be time-dependent in the Lagrangian frame. 

Depending on the mechanism that provides the effective viscosity of the disk, we will in the following describe two models for the evolution of the gas ring. In our first model, we will assume that the viscosity is provided by turbulence and/or magnetic fields, leading to a characteristic surface density profile of $\Sigma\propto R^{-1}$ in the stationary case, as well as a second model which assumes that the viscosity is provided by gravitational stresses. The latter requires a steeper density profile.

\subsection{Generic disk model}\label{generic}
As outlined above, we consider a radial annulus of mass $dM$ and surface density $\Sigma$ at radius $R$ in an axisymmetric primordial disk assumed to be stationary in the Eulerian frame. As a result, the accretion rate $\dot{M}$ is independent of time and position within the disk. The sound speed in the annulus is given as\begin{equation}
c_s=\sqrt{\gamma k_B T/m},
\end{equation}
with $k_B$ the Boltzmann constant, $T$ the gas temperature and $m$ the mean molecular mass. {In the following, we need to construct a dynamical model for the evolution of the annulus, in particular the radius $R$ and the surface density $\Sigma$. The latter is necessarily approximate, as we are not considering a fully Keplerian disk, but rather a situation where the disk is also stabilized by turbulent pressure, as reflected in the rather thick disks forming in primordial environments \citep[see e.g.][]{Turk10, Clark11, Greif12, Latif13PopIII}.}

{In the generic disk model, we will initially consider a situation with $\Sigma\propto R^{-1}$, as in the case of a so-called Mestel disk. The latter corresponds to a steady-state solution for a strongly self-gravitating disk with constant sound speed $c_s$ and a constant Toomre-Q parameter \citep{Toomre64}}
\begin{equation}
Q\sim\frac{c_s\Omega}{\pi G \Sigma}\label{Toomre}
\end{equation}
{of the order 1 \citep{Bertin97, Lodato07}. Strictly speaking, a Mestel disk further extends to infinity, while we are here considering disks of a finite size. The latter implies that also the rotation curve derived from the Mestel disk holds only in an approximate way. While the assumptions employed here are certainly an approximation, for instance the surface density distribution in realistic disks is often similar to the $R^{-1}$ profile \citep[e.g.][]{Beckwith90}, and we therefore employ it here in our approximate model. The evolution equation of the disk surface density in an axisymmetric model is now given as \citep{Lodato07}}
\begin{equation}
\frac{\partial \Sigma}{\partial t}=-\frac{1}{R}\frac{\partial }{\partial R}\left[ \frac{1}{(R^2\Omega)'}\frac{\partial}{\partial R}(\nu \Sigma R^3 \Omega') \right],
\end{equation}
{where the prime $'$ denotes the derivative with respect to $R$. Under the assumption that $\nu$ and $\Omega$ are independent of $\Sigma$, which certainly is an idealization, it can be shown that the disk surface density evolves on a timescale of the order of the viscous timescale, which is given as}
\begin{equation}
t_{vis}=\frac{R^2}{\nu},
\end{equation}
{where $\nu$ the effective viscosity of the gas. Considering the disk annulus in our model, the evolution of the surface density is thus approximately given as} \begin{equation}
\dot{\Sigma}\sim\frac{\Sigma}{t_{vis}}.\label{sigma}
\end{equation}
{In the case of a Keplerian disk, this expression becomes accurate by including an additional factor of $3/2$. We will however consider more generic situations, where the disk is also stabilized by turbulence, and not necessarily dominated by a central source. As we assume here a disk profile with $\Sigma\propto R^{-1}$, the radius of the annulus must follow a similar evolution equation given as}\begin{equation}
\dot{R}=-\frac{R}{t_{vis}}.\label{radius}
\end{equation}
{For a stationary thin disk, the continuity equation becomes \citep{Lodato07}}\begin{equation}
\dot{M}=\left|\frac{d\ln\Omega}{d\ln R}  \right| 2\pi\nu \Sigma.
\end{equation}
{For a generic disk model, the factor $|d\ln\Omega/d\ln R|$ will usually be of order $1$, and becomes equal to $3/2$ for a Keplerian thin disk. We therefore employ the approximate expression}
\begin{equation}
\dot{M}=3\pi\nu\Sigma.\label{eqMdot}
\end{equation}
{In our model, we assume that the constant accretion rate $\dot{M}$ is known, and the evolution of the disk surface density follows from Eq.~\ref{sigma}. We can therefore employ Eq.~\ref{eqMdot} to solve for the disk viscosity required to maintain the accretion rate, yielding}
\begin{equation}
\nu=\frac{\dot{M}}{3\pi \Sigma}.\label{nu}
\end{equation}
While this identity follows from the assumption of stationarity and axisymmetry, such conditions can only be achieved if a physical mechanism is present to provide the effective viscosity $\nu$. {We note that the formulation employed here is clearly approximate, and the evolution in the disk can depend both on the source of the viscosity, which can be due to turbulence and magnetic fields \citep[see e.g.][]{Balbus99, Hawley00} or the presence of gravitational instabilities \citep{Toomre64}. We assume here that the Toomre Q parameter is at least initially of order $1$, implying that self-gravity is relevant when the annulus forms.   If the gravitational force is balanced by the centrifugal force, the Keplerian angular velocity in a disk dominated by self-gravity is given as $\Omega_K=\sqrt{2\pi G \Sigma/R}$ \citep{Lodato07}. In a realistic disk, we however expect it to be partly supported by thermal and turbulent pressure, as also demonstrated via  numerical simulations \citep[e.g.][]{LatifBH, Latif13PopIII}. For the disk-dominated case, we therefore adopt}
\begin{equation}
\Omega_{\rm disk}=\epsilon_K\sqrt{\frac{2\pi G \Sigma}{R}},\label{Omegadisk}
\end{equation}
{where $\epsilon_K$ describes the deviation from a fully rotationally supported disk and $G$ denotes the gravitational constant. We will in the following assume a generic value $\epsilon_K=50\%$, as primordial disks typically have scale heights with $H/R$ between $0.3-1\%$ \citep{Turk10, Clark11, Greif12, LatifBH, Latif13PopIII}. As we assume that $\Sigma\propto R^{-1}$, the latter implies that the Toomre Q parameter remains $\sim1$ if the sound speed $c_s$ remains approximately constant. We note that our surface density profile further implies a non-zero enclosed mass in the innermost region of our disk. For the disk models considered here, the maximum mass within the central $0.1$~AU is $0.003$~M$_\odot$, which is expected not to be relevant for the dynamical evolution.}

{If the disk is dominated by the central source, we similarly assume that the angular velocity is given as a fraction $\epsilon_K$ of the Keplerian rotation, implying that} \begin{equation}
\Omega_{\rm source}= \epsilon_K \sqrt{\frac{GM_*}{R^3}},
\end{equation}
{with $M_*$ the mass of the central source. In this case, it  follows that $Q\propto c_s\Omega/\Sigma\propto R^{-3/2}/R^{-1}\propto R^{-1/2}$, i.e. $Q$ is increasing with decreasing radius, and the interior of the disk becomes gravitationally stable. As long as a sufficient viscosity is provided by turbulence and magnetic fields, the latter is not a problem for the assumed stationarity of the disk. In the next subsection, we will  consider the possibility that the viscosity is provided by self-gravitational instabilities, which will require a steeper relation between $\Sigma$ and $R$.}
We now calculate the disk height as\begin{equation}
H=\frac{c_s}{\Omega}.
\end{equation} 
The latter expression holds for gravitationally stable or marginally unstable disks \citep{Lodato07}, and is thus valid both for $Q\sim1$ and $Q>1$. The number density in the mid plane of the disk follows as\begin{equation}
n=\frac{\Sigma}{2Hm},
\end{equation}
and the mass density is given as $\rho=nm$. With these quantities, we can also evaluate the viscous heating rate. The latter is given as \citep{Lodato07, Ferrara13}
\begin{equation}
Q_+ = \nu \Sigma (R \Omega')^2.
\label{eq5}
\end{equation}
The viscous heating rate therefore generally depends on the rotational profile. In case of rotation around a central source, the viscous heating rate can be evaluated as
\begin{equation}
Q_{\rm+,source} = \frac{9}{4} \nu \Sigma \Omega_{\rm source}^2.
\label{eq6}
\end{equation}
For a disk dominated by self-gravity, on the other hand, we have $\Omega\propto\sqrt{\Sigma/R}\propto R^{-1}$, implying that
\begin{equation}
Q_{\rm+,disk} = \nu \Sigma \Omega_{\rm disk}^2.
\label{eq7}
\end{equation}
{In this framework, we do not impose thermal equilibrium, but rather follow the non-equilibrum evolution of the gas temperature in our annulus using the detailed chemical model described in section~\ref{chem}. The corresponding evolution equation is given as \citep{Grassi14}}\begin{equation}
\frac{dT}{dt}=(\gamma-1)\frac{\Gamma-\Lambda}{k_B n},
\end{equation}
{where $n$ is the number density of the gas following from disk surface density, disk height and chemical composition, $\Gamma$ is the total heating rate including viscous heating, and $\Lambda$ the total cooling rate. A further description of these expressions is given in section~\ref{chem}. While following the non-equilibrium evolution of the system is in principle more accurate than the assumption of thermal equilibrium, we note that the resulting state generally leads to a situation close to thermal equilibrium, i.e. with $\Gamma\sim\Lambda$.}

\subsection{Self-regulated disk model}\label{self-regulated}
As already shown above, a disk dominated by a central source is gravitationally stable in the interior if the evolution of the annulus is dictated by Eqs.~\ref{sigma} and \ref{radius}, implying $\Sigma\propto R^{-1}$. Such a scenario can thus only be maintained if a sufficient viscosity is provided by turbulence and magnetic fields, as assumed in the generic disk model. However, it is conceivable that gravitational instabilities are required to transport the angular momentum, and they were also shown in previous studies to provide the stronger contribution \citep[e.g.][]{Fromang04}. While the reality may lie somewhere in between, we will explore here the extreme case where self-gravity is necessary to provide the viscosity, requiring that $Q=1$. 

In this model, we assume that the disk is gravitationally unstable on scales larger than the current radius $R$, implying that gas is transported onto the annulus from larger scales, and the surface density of the annulus keeps increasing as in Eq.~\ref{sigma}. The radius $R$ however remains constant as long as $Q>1$, implying $\dot{R}=0$, and only evolves as in Eq.~\ref{radius} if $Q\leq1$. This condition will effectively alter the relation between $\Sigma$ and $R$, and increase the surface density until a state of marginal stability is reached. As a result, one obtains a higher surface density and number density on a given scale, and the ability of the gas to cool increases compared to the generic disk model outlined above.

\subsection{Chemical model}\label{chem}
To study the chemical evolution for the disk models outlined above, we employ the publicly available chemistry package \textsc{krome} developed by \citet{Grassi14}. In this framework, we adopt a network similar to the one described by \citet{Latif15X}, including the species H, H$^+$, H$^-$, H$_2^+$, H$_2$, He, He$^+$, He$^{++}$ and e$^-$. The photo-rates are discarded here, due to the efficiency of self-shielding at high densities. The network includes the currently most accurate rate coefficient for the 3-body H$_2$ formation  derived by \citet{Forrey13} from on a quantum-mechanical calculation \citep[see also discussion by][]{BovinoH2}. 

The cooling functions in \textsc{krome} include the atomic line cooling, recombination cooling and Bremsstrahlung cooling as described by \citet{Cen92}, the H$_2$ cooling function derived by \citet{Glover08} as well as the H$_2$ formation heating and cooling as described by \citet{Omukai00}. Continuum cooling processes are treated as described by \citet{Omukai00} with the continuum opacities derived by \citet{Lenzuni91}, employing a new fit provided by \citet{Grassi14}. The \citet{Lenzuni91} opacities include all relevant continuum processes, specifically the bound-free absorption by H and H$^-$, free-free absorption by H, H$^-$, H$_2$, H$_2^-$, H$_2^+$, H$_3^+$, He and He$^-$, Rayleigh scattering by H, H$_2$ and He and collisionally-induced absorption by H$_2$. The viscous heating is included as described in section~\ref{theory}.

The chemical model described here will be called within each dynamical timestep of the disk model outlined above to follow the chemical and thermal evolution. As a result, we obtain the evolution of the chemistry while the gas is moving inward. The latter corresponds to a time sequence in the Lagrangian frame of the annulus, but also describes the structure of the stationary disk in the Eulerian frame. In the following, we will typically show the results as a function of radius to illustrate the resulting structure of the disk.

\section{Chemical evolution for an initially molecular gas}\label{molecular}
In the following, we present the results for the chemical evolution in primordial viscous disks for the scenarios outlined above. For the chemical initial conditions, we consider an initially molecular gas. We adopt a helium fraction of $7.75\%$ with respect to the total number of nuclei. The typical densities at the beginning of the calculation are of the order $\sim10^9-10^{11}$~cm$^{-3}$. For a conservative assessment for the impact of viscous heating, we assume here that the gas is very close to fully molecular, with an atomic hydrogen fraction of only $10^{-6}$. For the electron/proton abundance, we adopt a generic value of $10^{-9}$, which can be expected at such densities \citep{Bovino13b}. The further species abundances like H$^-$ and H$_2^+$ are initially set to zero, but almost instantaneously reach their equilibrium values during the chemical evolution. We also checked that the results do not strongly depend on these assumptions.

In the following, we will present the chemical evolution for such an initially molecular gas for the case of a generic disk model, both within and without a central source, as well as for a self-regulated disk model, for which we predominantly focus on the case with a central source. 

\subsection{Results for a generic disk model with no central source}\label{Mestelnosource}

For a generic disk model with no central source, we explore a range of scenarios with different accretion rates, varying between $10^{-3}-10^0$~M$_\odot$~yr$^{-1}$, ranging from typical Pop.~III star formation \citep{Abel02, Bromm03} to the conditions in atomic cooling haloes \citep{Latif13BHmass, Regan14}. We assume the initial annulus to be at a radius $R=1000$~AU, with a characteristic temperature of $\sim300$~K. The surface densities of the disk are determined via the model of \citet{Latif15}, implying an initial condition with $Q=1$. The details of the models are given in Table~\ref{tab:nosource}.

\begin{table}[htp]
\begin{center}
\begin{tabular}{c|c|c|c|c}
Model & $\dot{M}$~[M$_\odot$~yr$^{-1}$] & $\Sigma$~[g~cm$^{-2}$] & $R$~[AU] & $T$~[K] \\ \hline
NS0 & $10^0$ & $100$ & $1000$ & $300$\\
NS1 & $10^{-1}$ & $50$ & $1000$ & $300$\\
NS2 & $10^{-2}$ & $25$ & $1000$ & $300$\\
NS3 & $10^{-3}$ & $12.5$ & $1000$ & $300$
\end{tabular}
\end{center}
\caption{Generic disk models with no central source and an initially molecular gas.}
\label{tab:nosource}
\end{table}%

\begin{figure}[h]
\begin{center}
\includegraphics[scale=0.7]{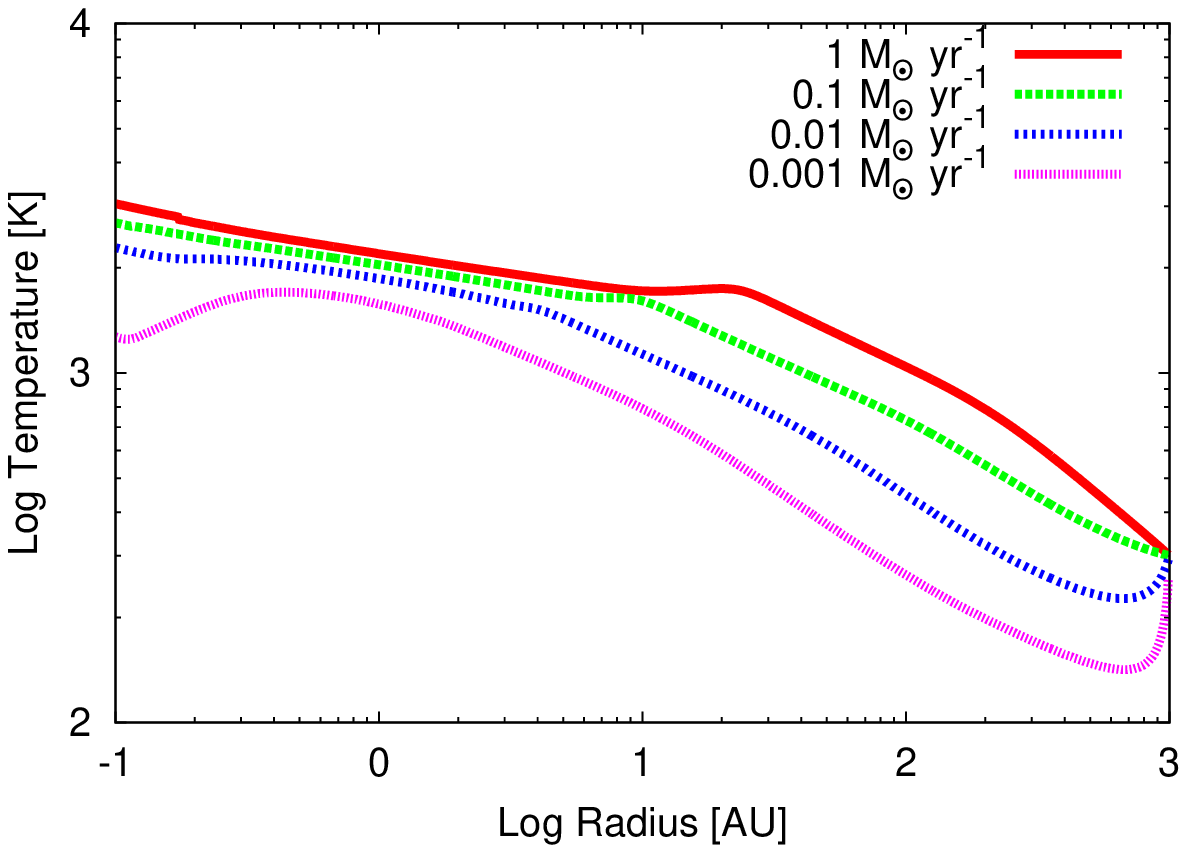}
\includegraphics[scale=0.7]{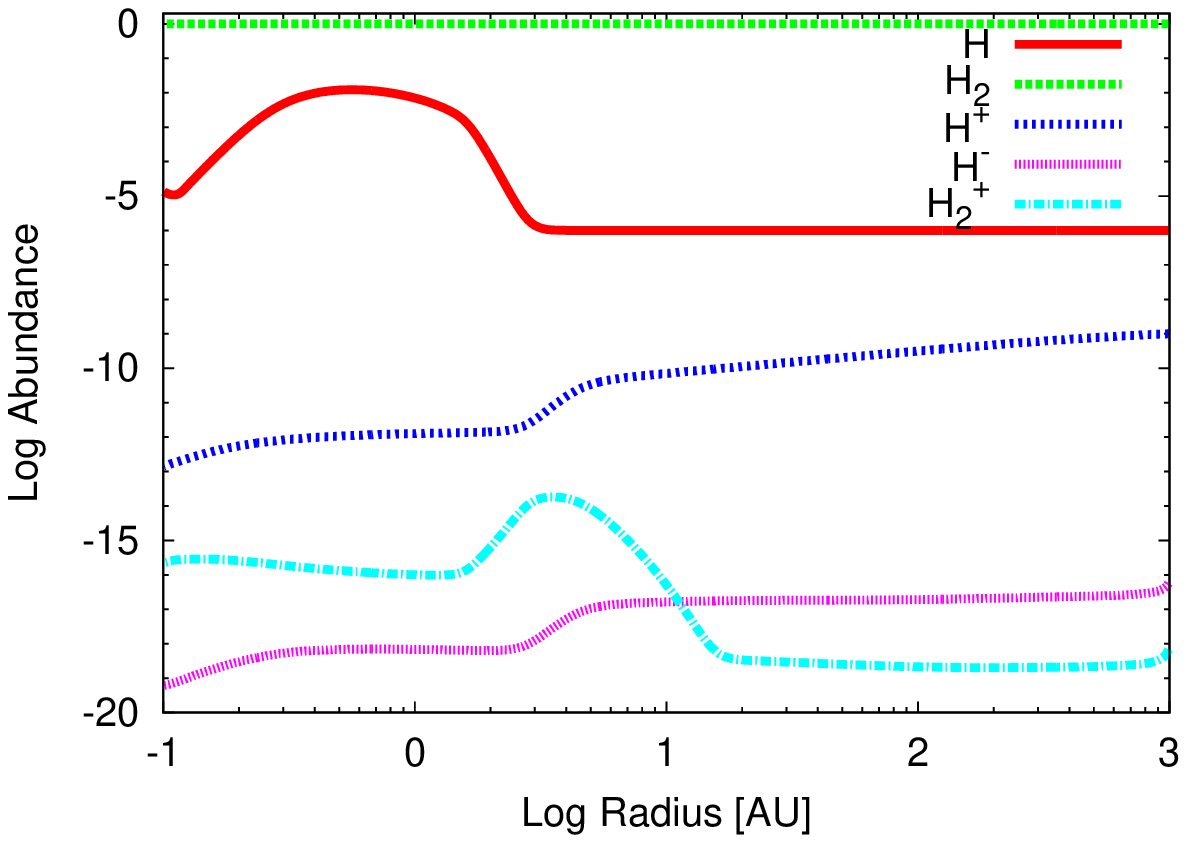}
\includegraphics[scale=0.7]{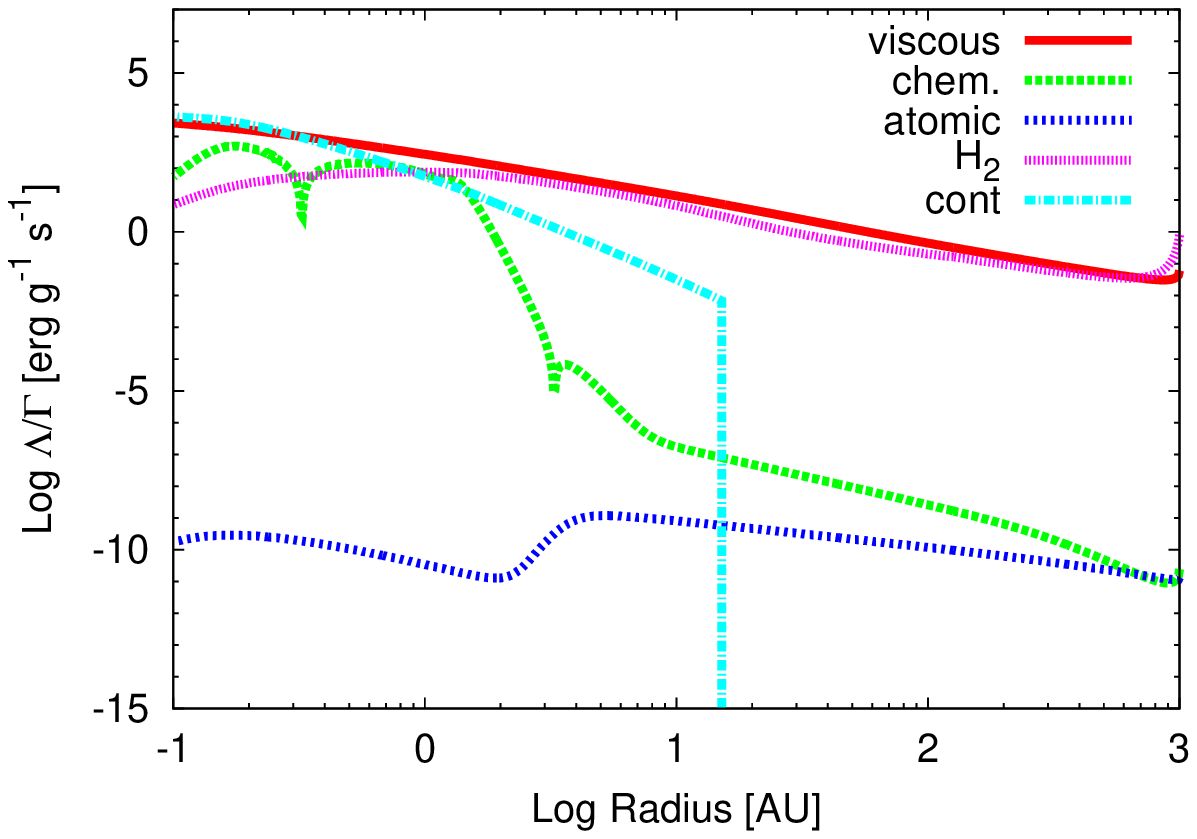}
\caption{Results for the generic disk models without central source and an initially molecular gas (Table~\ref{tab:nosource}). Top panel: Gas temperature vs radius in case of a generic disk model without central source, assuming different accretion rates. Mid panel: Abundances of H, H$_2$, H$^+$, H$_2^+$, H$^-$ vs radius for the reference case NS3 (disk with no source, accretion rate $10^{-3}$~M$_\odot$~yr$^{-1}$). Bottom panel: Heating and cooling contributions vs radius for the reference case NS3 (disk with no source, accretion rate $10^{-3}$~M$_\odot$~yr$^{-1}$). We refer to Table~\ref{tab:cool} for the heating/cooling contributions defined in the legend.}
\label{nosource}
\end{center}
\end{figure}

As already shown in section~\ref{generic}, the surface density scales as $R^{-1}$ under these conditions, and the surface densities reach peak values between $10^5$ and $10^6$~g~cm$^{-2}$ at $R=1$~AU. The results for this case are given in Fig.~\ref{nosource}. We find that the temperature generally increases towards the interior as a result of viscous heating $Q_+\propto\nu\Sigma\Omega_{\rm disk}^2$ (top panel). As $\nu\propto\Sigma^{-1}$ (Eq. \ref{nu}), the latter essentially scales as $\Omega_{\rm disk}^2\propto \Sigma/R\propto R^{-2}$, implying a steep increase as a function of radius, which  is only partly compensated by the increasing number density of the gas. In the regime considered here, we note that H$_2$ is already thermalized, so that the cooling rate only scales linearly with $n_{\rm H_2}$, and optically depth effects are increasing with density. The gas temperature therefore increases towards values between $1000-2000$~K, and then rises more gradually.

This behavior is generally found in all cases considered here, with minor deviations for an accretion rate of $0.001$~M$_\odot$~yr$^{-1}$, where the gas temperature slightly decreases again near $0.1$~AU, as viscous heating is reduced in this case and the collisionally-induced emission (CIE) of molecular hydrogen simultaneously becomes relevant. We further note that the temperature of the gas increases with the accretion rate. This is due to the proportional increase in viscous heating, which is only partly compensated by the increased number densities and the ability of the gas to cool. The viscous heating  has therefore a relevant impact on the thermal evolution of the gas, even though the differences in the temperature are here smaller than a factor of~$2$.

The chemical abundances for a characteristic case with an accretion rate of $0.001$~M$_\odot$~yr$^{-1}$ (model NS3) are shown in Fig.~\ref{nosource} (mid panel). We find that the gas is fully molecular during the entire evolution. The atomic hydrogen abundance is initially constant, temporarily increases towards $10^{-2}$ around $R\sim1$~AU as a result of the temperature increase, and decreases again in the interior, reflecting the slightly decreasing temperature. Similarly, the proton and electron abundance decreases towards the interior as a result of the increasing densities. 

The resulting contributions to heating and cooling are given in Fig.~\ref{nosource} (bottom panel), again for the case of NS3. We note  that the viscous heating rate is almost balanced by the molecular hydrogen cooling throughout most of the evolution between $1000$ and $1$~AU, and subsequently the continuum cooling dominates in the interior. Within $\sim1$~AU, we also find strong chemical heating and cooling, which are however tightly balanced, as the chemistry is close to its equilibrium value. During most of the evolution, the disk  remains in the molecular cooling regime, without any strong transitions in the gas temperature.

\subsection{Results for a generic disk model with central source}

Now, we consider the case of a generic disk model with a central source. For this purpose, we consider stellar masses  from $10-10^4$~M$_\odot$,  ranging from typical Pop.~III  to supermassive primordial stars, and accretion rates of $10^{-3}-10^0$~M$_\odot$~yr$^{-1}$. We start following the evolution at $R=1000$~AU, taking the initial surface density and gas temperature from the disk model of \citet{Latif15}. The details of the models  are given in Table~\ref{tab:source}.

\begin{table}[htp]
\begin{center}
\begin{tabular}{c|c|c|c|c}
Model & $M_*$~[M$_\odot$] &  $\dot{M}$~[M$_\odot$~yr$^{-1}$] & $\Sigma$~[g~cm$^{-2}$] & $T$~[K] \\ \hline
S1A1 & $10^1$ & $10^{-1}$ & $10$ & $300$\\
S1A3 & $10^1$ & $10^{-3}$ & $10$ & $300$\\
S2A1 & $10^2$ &$10^{-1}$ & $40$ & $500$\\
S2A3 & $10^2$ & $10^{-3}$ & $40$ & $500$\\
S4A0 & $10^4$ & $10^0$ & $790$ & $1900$\\
S4A1 & $10^4$ & $10^{-1}$ & $630$ & $1100$\\
\end{tabular}
\end{center}
\caption{Models for a generic disk model with central source and an initially molecular gas.}
\label{tab:source}
\end{table}%

\begin{figure}[h]
\begin{center}
\includegraphics[scale=0.7]{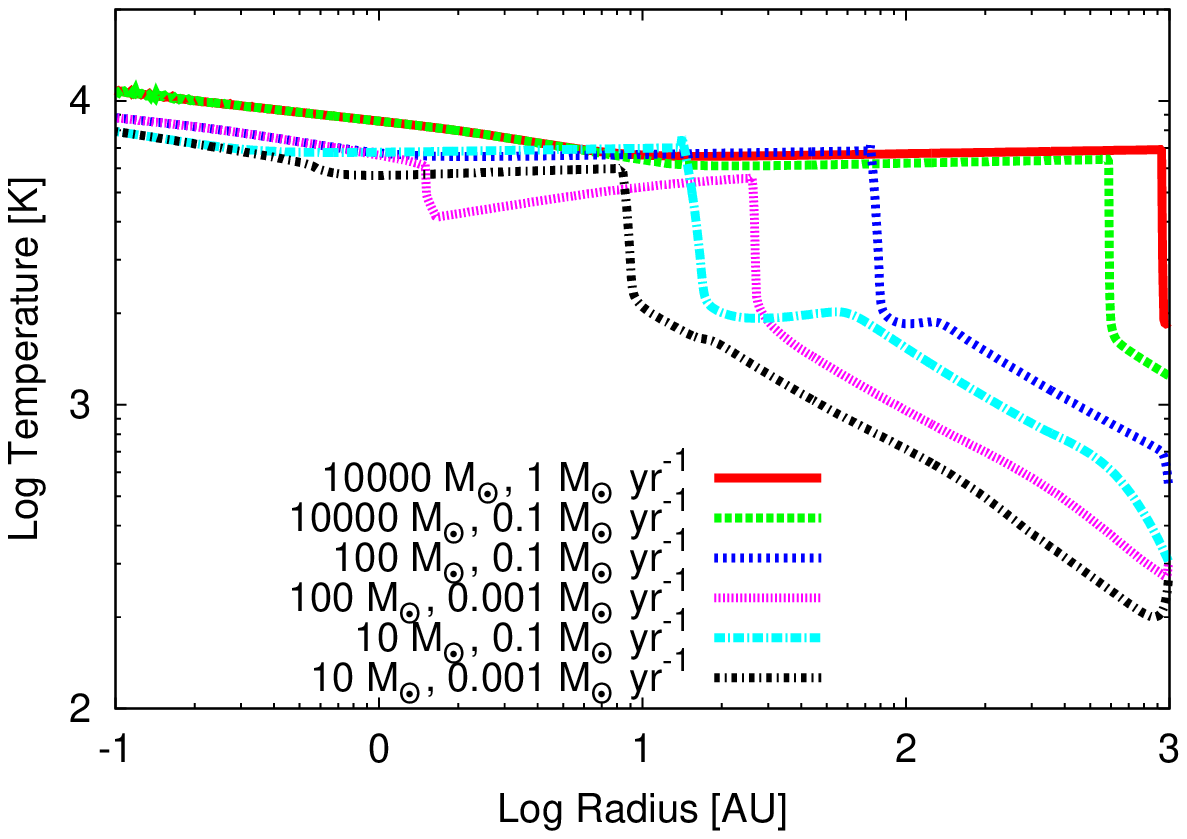}
\includegraphics[scale=0.7]{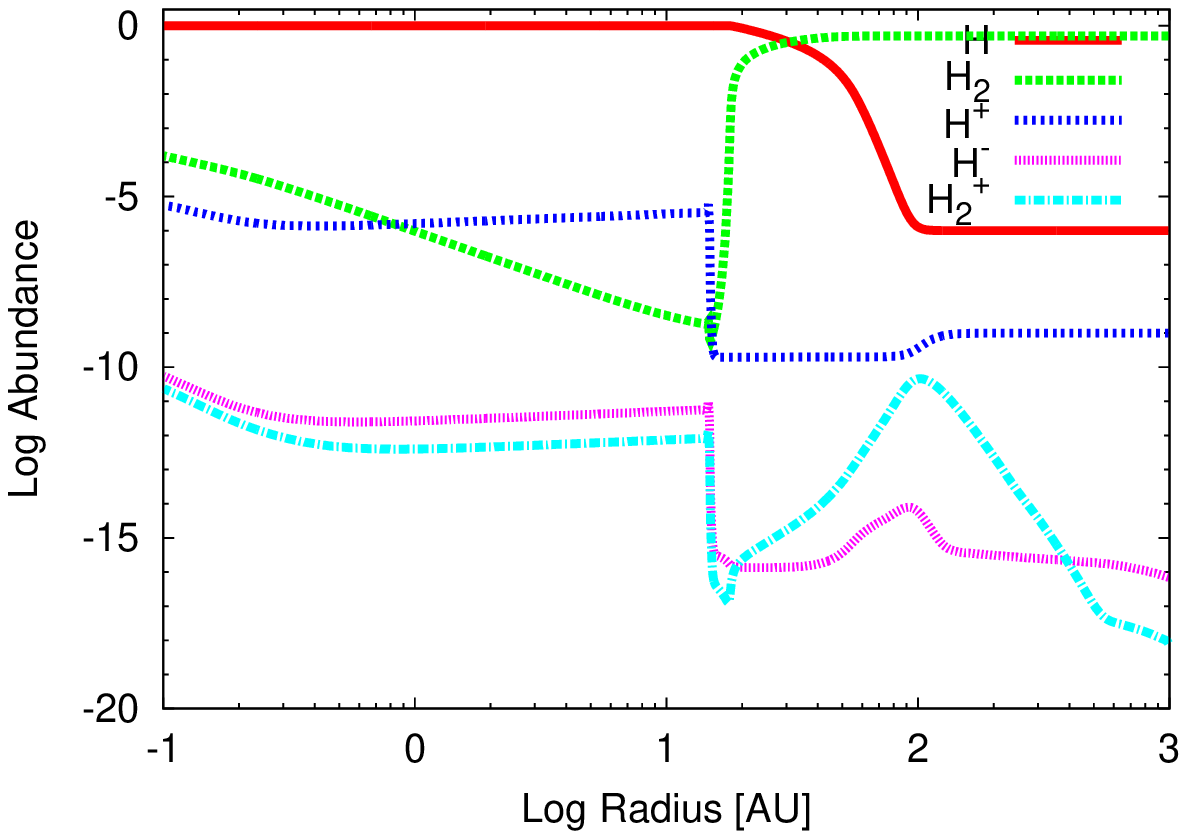}
\includegraphics[scale=0.7]{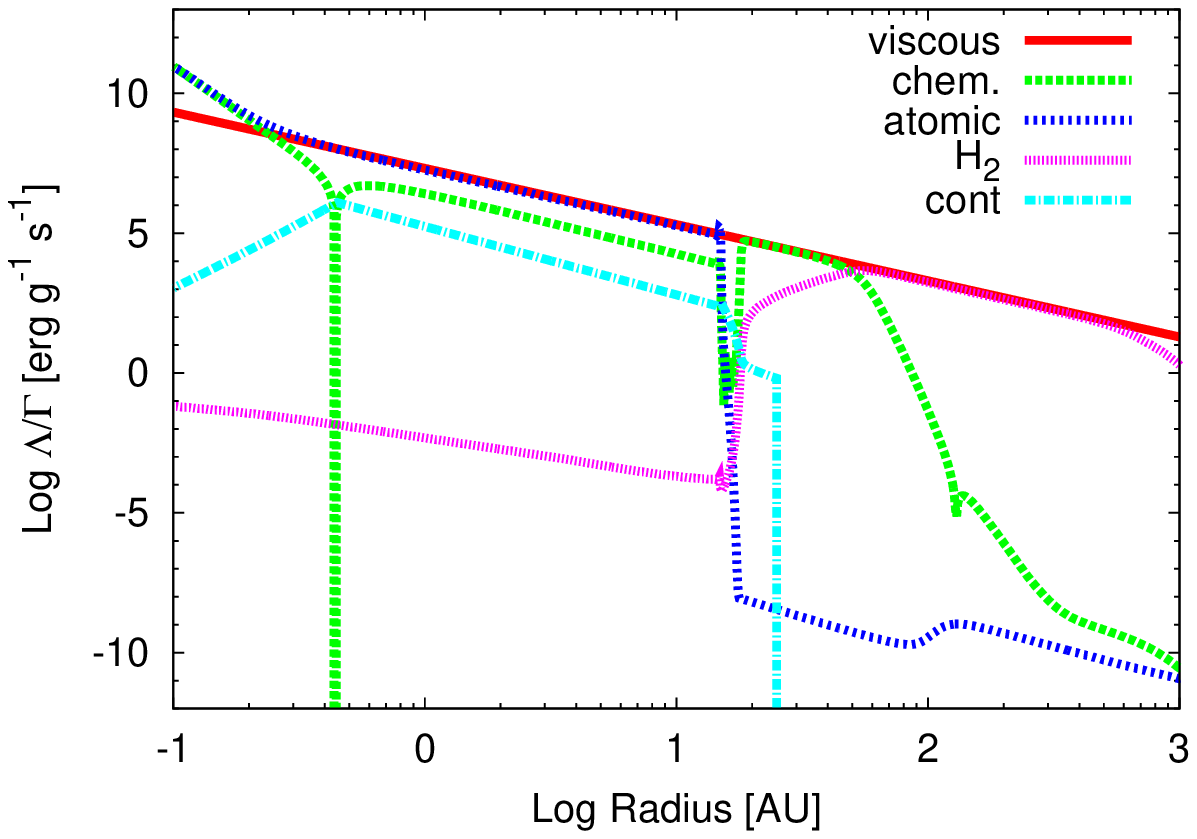}
\caption{Results for the generic disk models with central source and an initially molecular gas (Table~\ref{tab:source}). The species abundances and heating/cooling rates are given for the S2A3 (generic disk model with central source of $100$~M$_\odot$ and accretion rate of $10^{-3}$~M$_\odot$~yr$^{-1}$). We refer to the caption of Fig.~\ref{nosource} for the description of the panels.}
\label{source}
\end{center}
\end{figure}

In this case, the surface density still scales as $R^{-1}$, reaching peak values of $10^5-10^7$~g~cm$^{-2}$. The temperature however changes more significantly during the evolution. For a characteristic case with a stellar mass of $10$~M$_\odot$ and an accretion rate of $10^{-3}$~M$_\odot$~yr$^{-1}$ (model S1A1), the temperature increases from $300$~K at $R=1000$~AU to $\sim2000$~K at $10$~AU, and then raises rapidly to $\sim7000$~K after only a minor change in radius as shown in Fig.~\ref{source} (top panel). This behavior reflects the steep dependence of the viscous heating rate with radius, scaling as $\Omega_{\rm source}^2\propto M_*/R^3$. The scaling relation is thus considerably steeper than for the case without a central source, and increases further with increasing stellar mass. Here, the temperature increases towards a value of $\sim2000$~K, where the collisional dissociation of H$_2$ becomes relevant. We note that the latter has an exponential dependence on the temperature, and therefore the molecular hydrogen becomes fully dissociated after the temperature increases further. At that point, the gas is no longer able to cool via molecular hydrogen, and therefore increases until the heating is balanced via atomic cooling. The temperature then remains high at $6000-8000$~K on scales below $10$~AU.

In general, the evolution is rather similar also for the other cases considered. We note  that the transition towards the atomic cooling regime occurs earlier for higher stellar masses and/or higher accretion rates, as both lead to an increase in the amount of viscous heating. For a $10^4$~M$_\odot$ star with an accretion rate of $1$~M$_\odot$~yr$^{-1}$, the viscous heating is particularly efficient, leading to an almost instantaneous transition already at $R\sim3000$~AU. This transition towards the atomic cooling regime may considerably increase the stability of the accretion disk in the environment of supermassive stars, as also described by \citet{Latif15}.

The chemical abundances for a representative case with a stellar mass of $100$~M$_\odot$ and an accretion rate of $10^{-3}$~M$_\odot$~yr$^{-1}$ (model S2A3) is given in Fig.~\ref{source} (mid panel), clearly showing the transition from the molecular to the atomic regime at $R\sim30$~AU. The latter is accompanied by a significant increase of the ionization fraction due to the higher temperature. We note that the molecular hydrogen abundance starts increasing again with density after the transition, but does not reach values where the H$_2$ cooling would be significant. In Fig.~\ref{source} (bottom panel), the heating and cooling rates are plotted for this reference case, confirming that the cooling is initially dominated by molecular hydrogen, and the atomic line cooling takes over after the transition. Within the central $0.3$~AU, we note that also chemical heating due to H$_2$ formation starts getting significant, but is still balanced by the atomic hydrogen cooling. The central temperature then raises to a value of $\sim9000$~K at $0.1$~AU.

\subsection{Results for self-regulated disks}
As already discussed in section~\ref{self-regulated}, it is conceivable that gravitational instabilities are required to provide a sufficiently large effective viscosity to transport the angular momentum, in particular in the regime of high accretion rates. We therefore consider the evolution in a self-regulated disk, starting with an initially molecular gas. We predominantly focus on the case with a central source. As we mentioned in section~\ref{generic}, the evolution in the case without a central source would be very similar to the generic disk model. We explore characteristic masses of the source of $1-10^3$~M$_\odot$, including typical Pop.~III stars and supermassive stars. Of course, even stronger effects can be expected in case of more massive stars.

\begin{table*}[htp]
\begin{tabular}{l|l|l}
Abbreviation & cooling and heating terms & reference\\ \hline
viscous & viscous heating & \citet{Lodato07} \\
chem.  & net chemical heating / cooling due & \citet{Omukai00}\\
chem. heat & chemical heating due to H$_2$ formation & \citet{Omukai00}\\
chem. cool & chemical cooling due to H$_2$ dissociation & \citet{Omukai00}\\
atomic & atomic line cooling, recombination cooling, Bremsstrahlung & \citet{Cen92}\\
H$_2$ & H$_2$ line cooling & \citet{Glover08}\\
cont & continuum cooling & \citet{Lenzuni91}\\ \hline
\end{tabular}
\caption{Cooling and heating functions listed under different abbreviations in Figs.~\ref{nosource}, \ref{source}, \ref{toomre}, \ref{hot} and \ref{hotq}.}
\label{tab:cool}
\end{table*}%

We focus here on typical accretion rates of $10^{-1}$~M$_\odot$~yr$^{-1}$, which is particularly relevant for the formation of supermassive stars in atomic cooling halos, and where the heating is particularly strong. We follow the evolution of the disk from a radius of $R=1000$~AU, and evaluate the initial surface density and gas temperature using the model of \citet{Latif15}. Due to the generally higher surface and number densities in this scenario, we restrict the evolution here to a minimum radius of $1$~AU. In fact, considering stellar evolution models for supermassive stars, their typical extend may even exceed such a scale \citep{Hosokawa13}. Overall, a summary of the scenarios considered here is given in Table~\ref{tab:self-regulated}.

\begin{table}[htp]
\begin{center}
\begin{tabular}{c|c|c|c|c}
Model & $M_*$~[$M_\odot$] & $\dot{M}$~[M$_\odot$~yr$^{-1}$] & $\Sigma$~[g~cm$^{-2}$]  & $T$~[K] \\ \hline
S1 & $1$ & $10^{-1}$ & $2.5$ & $300$ \\
S2 & $10$ & $10^{-1}$ & $10$ & $400$\\
S3 & $10^2$ & $10^{-1}$ & $40$ & $500$ \\
S4 & $10^3$ & $10^{-1}$ & $160$ & $800$ 
\end{tabular}
\end{center}
\caption{Models for a self-regulated disk with initially molecular gas.}
\label{tab:self-regulated}
\end{table}%

\begin{figure}[h]
\begin{center}
\includegraphics[scale=0.63]{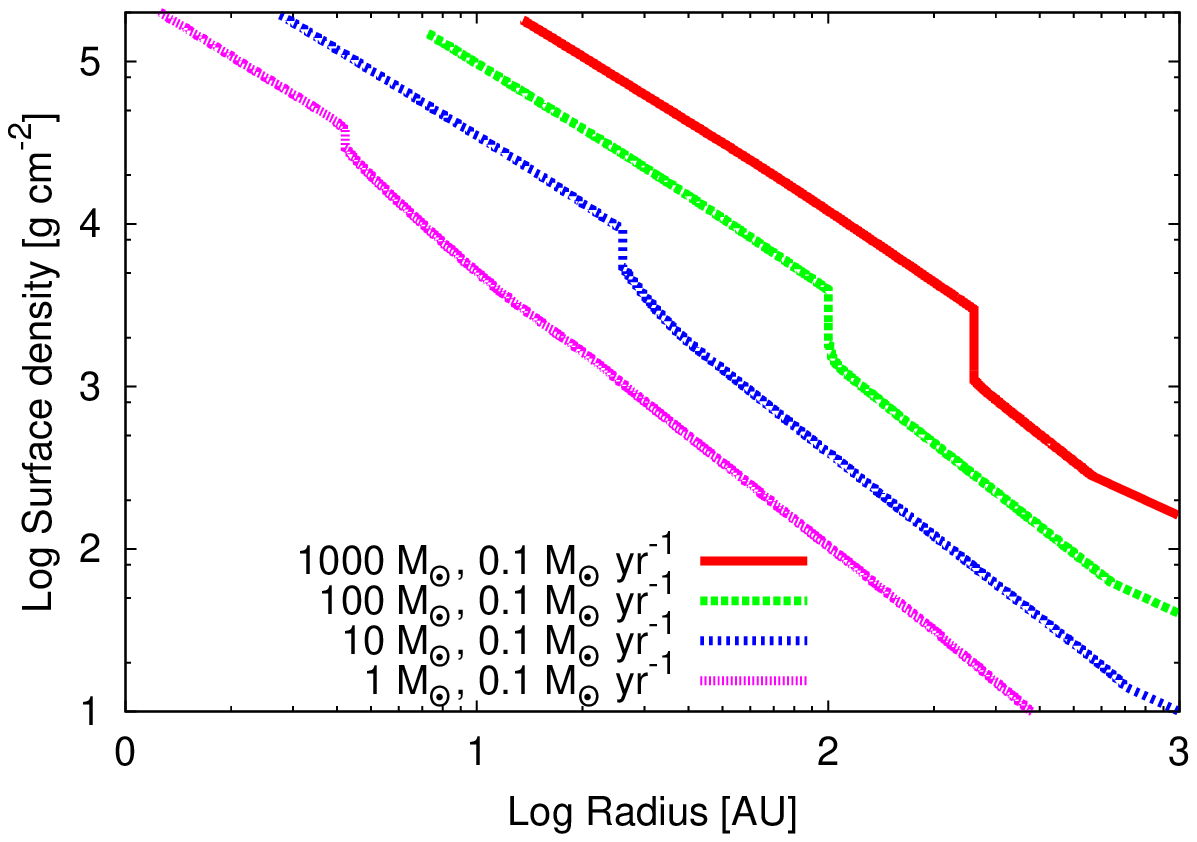}
\includegraphics[scale=0.63]{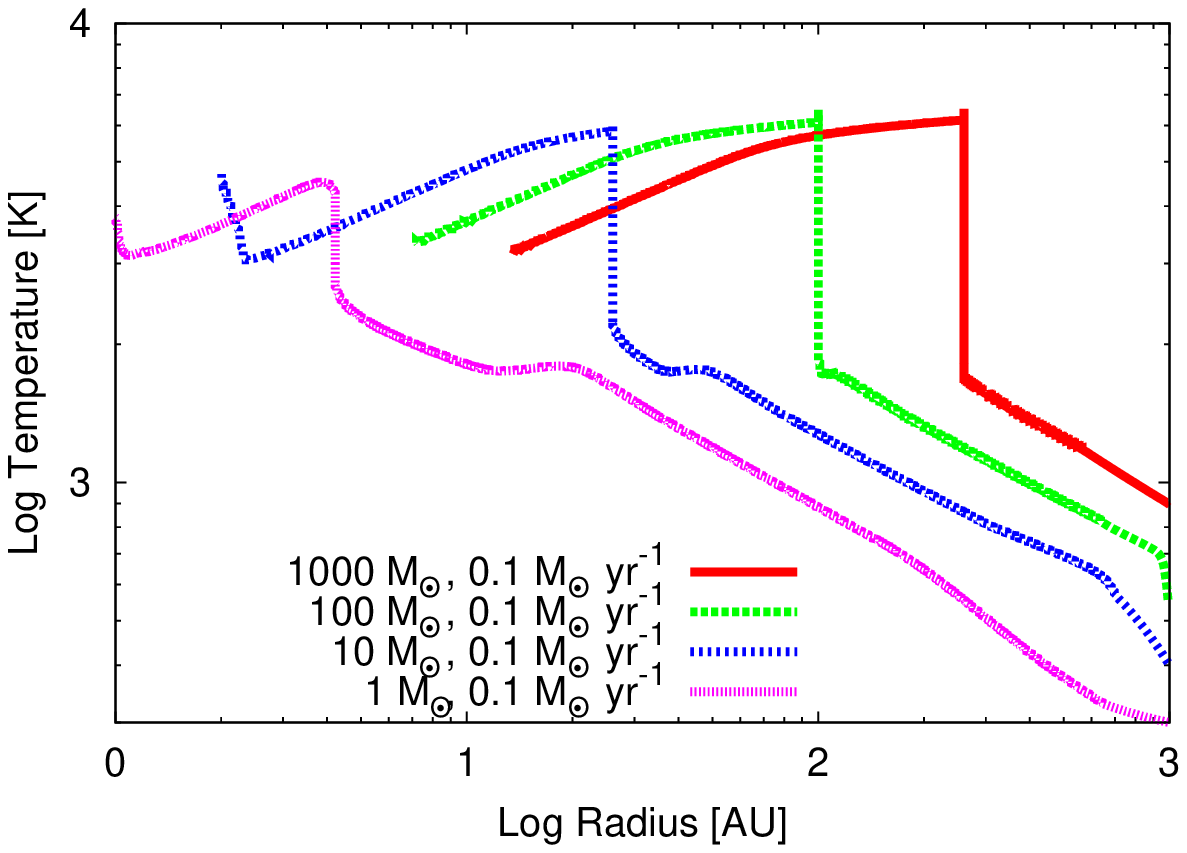}
\includegraphics[scale=0.63]{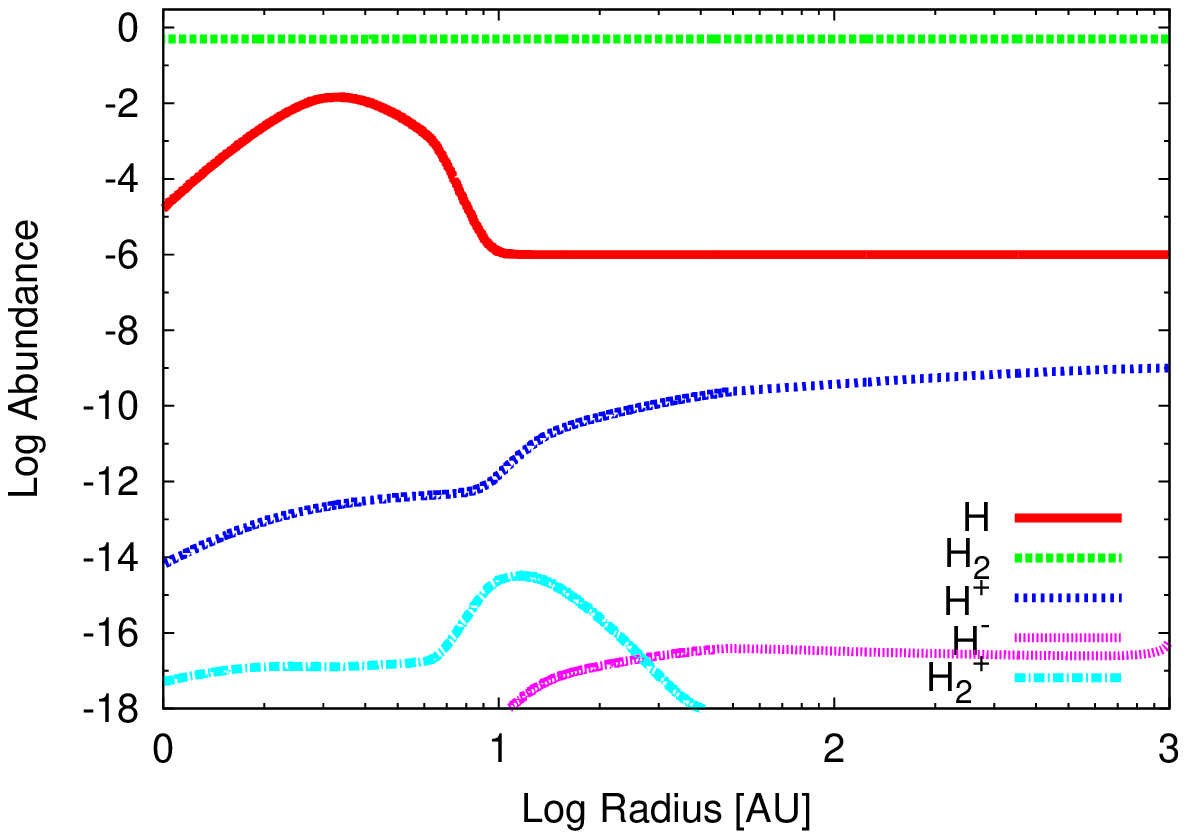}
\includegraphics[scale=0.63]{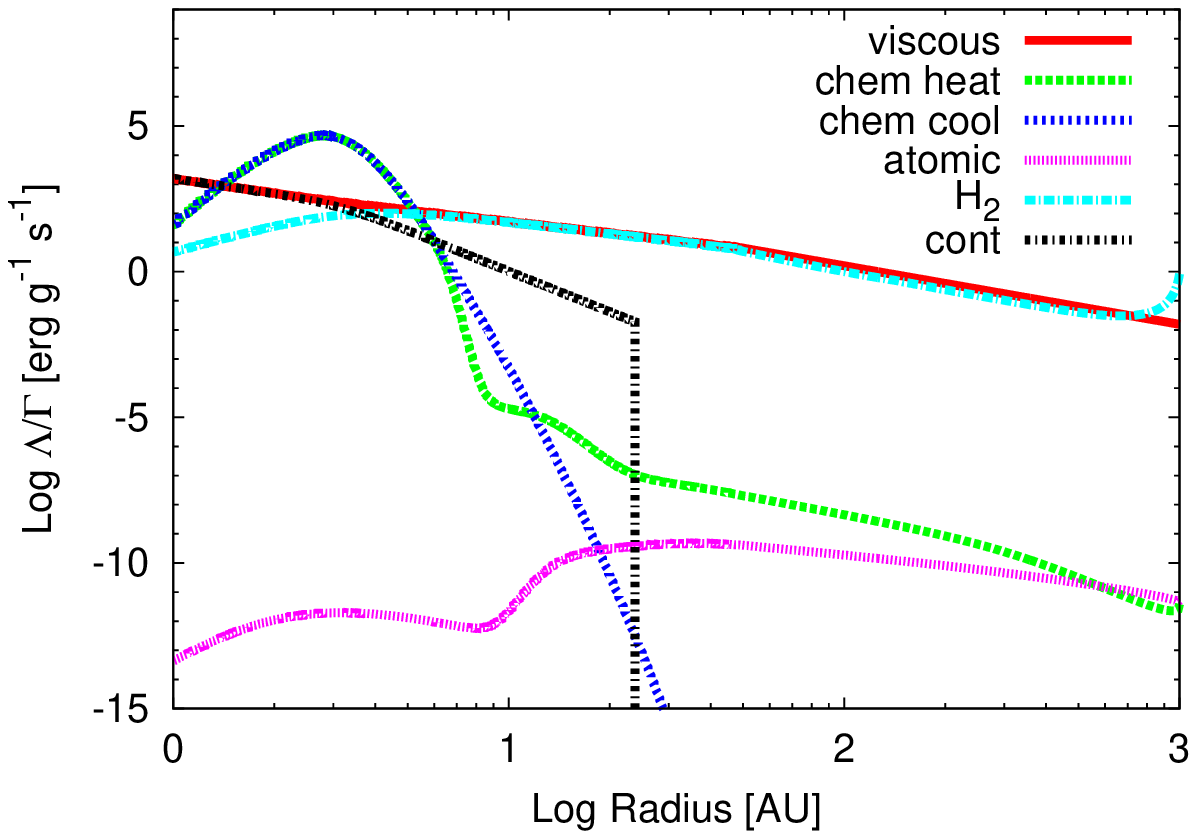}
\caption{Results for the self-regulated disk models  with an initially molecular gas (Table~\ref{tab:self-regulated}), plus species abundances and heating/cooling rates for the reference case S1 (self-regulated disk with central source of $1$~M$_\odot$ and an accretion rate of $10^{-1}$~M$_\odot$~yr$^{-1}$). The top panel shows the gas surface density as a function of radius. We refer to the caption of Fig.~\ref{source} for the description of the other panels. }
\label{toomre}
\end{center}
\end{figure}

For the self-gravitating disk models, due to the requirement of $Q=1$, the resulting relation between surface density $\Sigma$ and radius $R$ is considerably steeper than for a generic disk model, requiring roughly a relation of $\Sigma\propto R^{-1.5}$. From the results given in Fig.~\ref{toomre} (top panel), it is evident that characteristic features occur when the temperature increases strongly, i.e. at the transition from the molecular to the atomic cooling regime. In this case, the higher gas temperature needs to be compensated with a higher gas surface density to ensure the condition that $Q=1$, thus temporarily steepening the relation between surface density and radius. 

For a characteristic case with $1$~M$_\odot$ and an accretion rate of $10^{-1}$~M$_\odot$~yr$^{-1}$ (model S1), the temperature increases first from $400$~K at $1000$~AU to $\sim1800$~K at $30$~AU, and subsequently increases more gradually, due to the collisional dissociation cooling of molecular hydrogen, towards temperatures above $2000$~K near $3$~AU, where H$_2$ becomes fully dissociated and the temperature increases to $\sim5000$~K (Fig.~\ref{toomre}, 2nd panel). A similar behavior is shown for higher stellar masses. This is however more extreme, implying that the transition to the atomic  regime occurs earlier, with a higher characteristic temperature. For instance in the case of a supermassive star with $10^3$~M$_\odot$, the transition to the atomic  regime occurs already at $300$~AU, with a peak temperature of $7000$~K.

For our reference model S1, the evolution of the species abundances is given in Fig.~\ref{toomre} (3rd panel), confirming the transition from the molecular to the atomic regime near $10$~AU as well as the subsequent build-up of molecular hydrogen due to the high densities of the gas. It therefore starts affecting the cooling again around densities of $10^{16}$~cm$^{-3}$. The viscous heating is initially balanced by the molecular cooling between $30$ and $1000$~AU (Fig.~\ref{toomre}, bottom panel). Between $30$ and $3$~AU, the main cooling channel is due to H$_2$ collisional dissociation cooling. In the central $3$~AU, the contributions of viscous heating, continuum cooling as well as chemical heating and cooling strongly balance each other, and molecular hydrogen line cooling starts playing some role due to the increasing H$_2$ abundance. However, it is also visible for instance in the case S2 that the H$_2$ abundance is collisionally dissociated, as the critical temperature for H$_2$ dissociation decreases with density and the disk evolves into a denser regime. We expect similar effects to occur in the scenarios S3 and S4, where the calculation becomes however numerically unstable at this point as a result of the stronger viscous heating.

\section{Chemical evolution for an initially atomic gas}\label{atomic}

Now, we consider a case where the gas in the disk is initially atomic. In this case, we adopt an initial molecular hydrogen abundance of $10^{-6}$ and an initial temperature of $10^4$~K. Due to the higher temperature, we also assume an increased ionization degree of $10^{-6}$. In the following, we present the chemical evolution for an initially atomic gas both in the case of a generic disk model, as well as for a self-regulated disk.

\subsection{Results for a generic disk model with and without central source}

For the case of a generic disk model, we consider scenarios with and without a central source, with source masses of $10-10^4$~M$_\odot$. We adopt here an accretion rate of $10^{-1}$~M$_\odot$~yr$^{-1}$ to explore a case where viscous heating is particularly relevant. We start  with a characteristic surface density of $10$~g~cm$^{-2}$ at $1000$~AU. We  note that these disks are not necessarily self-gravitating, in particular in the presence of a central source, and the viscosity to build up the disk thus needs to be provided by turbulence or (magneto-)hyrodynamical instabilities. A summary of the explored models is given in Table~\ref{tab:Mestel}. 

\begin{table}[htp]
\begin{center}
\begin{tabular}{c|c|c|c|c}
Model & $M_*$~[$M_\odot$] & $\dot{M}$~[M$_\odot$~yr$^{-1}$] & $\Sigma$~[g~cm$^{-2}$] & $T$~[K] \\ \hline
H0 & $0$ & $10^{-1}$ & $10$  & $10^4$\\
H1 & $10$ & $10^{-1}$ & $10$ & $10^4$\\
H2 & $100$ & $10^{-1}$ & $10$ & $10^4$\\
H3 & $10^4$ & $10^{-1}$ & $10$ & $10^4$ 
\end{tabular}
\end{center}
\caption{Models for a generic disk model with an initially atomic gas.}
\label{tab:Mestel}
\end{table}%

\begin{figure}[h]
\begin{center}
\includegraphics[scale=0.65]{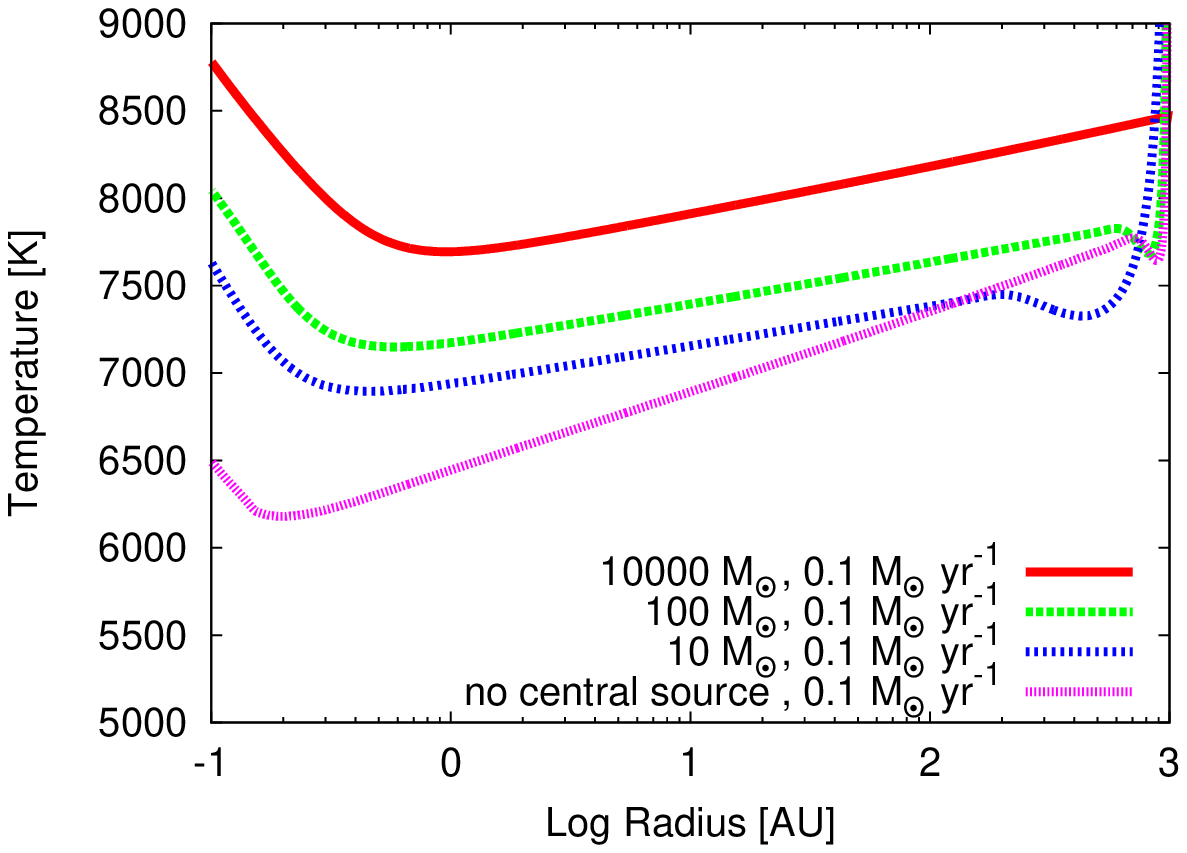}
\includegraphics[scale=0.65]{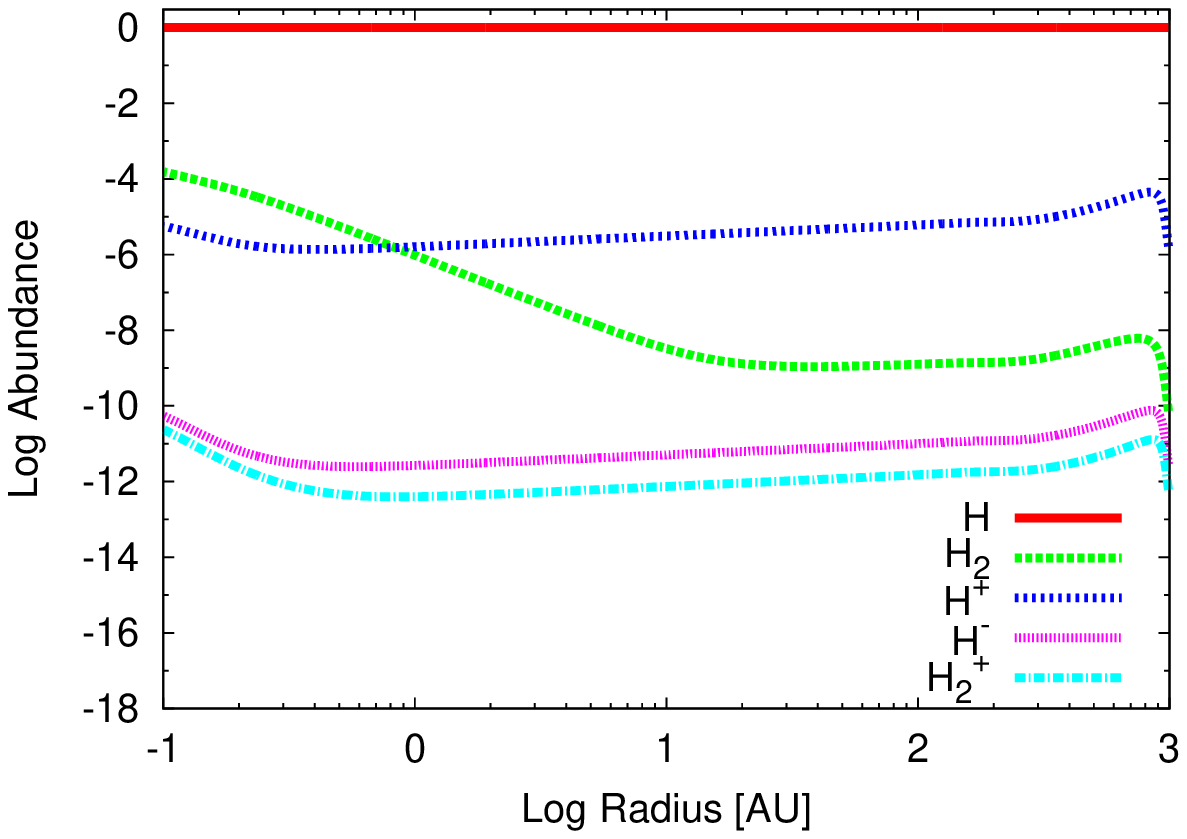}
\includegraphics[scale=0.65]{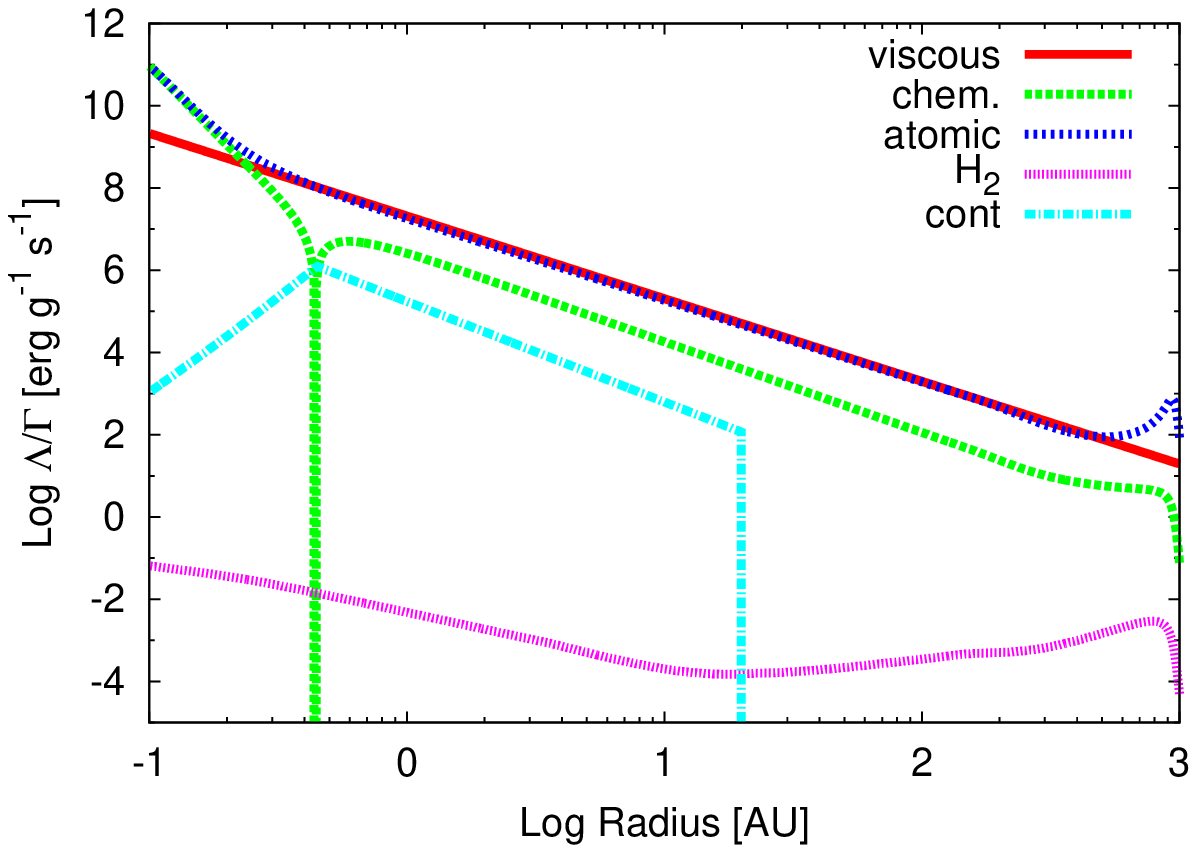}
\caption{Results for the generic disk model with an initially atomic gas (Table~\ref{tab:Mestel}). The species abundances and heating/cooling rates are given for the H1 (generic disk model with central source of $10$~M$_\odot$ and an accretion rate of $10^{-1}$~M$_\odot$~yr$^{-1}$). We refer to the caption of Fig.~\ref{nosource} for the description of the panels.}
\label{hot}
\end{center}
\end{figure}

The evolution of the gas temperature as a function of radius is given in Fig.~\ref{hot} (top panel). The temperature remains high throughout the entire evolution, from $R=1000$~AU to $R=0.1$~AU, but decreases slightly from initially $8000-9000$~K to $6000-9000$~K in the interior. We note that the tem-perature  slightly increases with central mass as a result of the enhanced viscous heating. The role of viscous heating is overall less pronounced  compared to the initially molecular regime, as also found by \citet{Ferrara13}. In fact, the temperature remains high even in the absence of a central source, though may somewhat increase in the presence of viscous heating.

For a characteristic case with a central source of $10$~M$_\odot$ and an accretion rate of $10^{-1}$~M$_\odot$~yr$^{-1}$ (model H1), the resulting species abundances are plotted in Fig.~\ref{hot} (mid panel). We note  that the gas is atomic in the entire regime, and H$_2$ remains strongly suppressed via collisional dissociation due to the high temperature. Within the central $10$~AU, the H$_2$ abundance starts increasing due to a mild decrease of the temperature, however leading only to an abundance of $10^{-4}$ at $0.1$~AU, which is insufficient to drive the cooling. The ionization degree only evolves very gradually as a function of radius, and is slightly decreasing due to the increasing densities. The viscous heating is balanced by the atomic cooling through almost the entire evolution (Fig.~\ref{hot}, bottom panel). Only in the very central $0.3$~AU, chemical heating via three-body H$_2$ formation becomes important as well, but is still balanced by the atomic line cooling. The molecular hydrogen cooling remains negligible throughout the entire evolution.

\subsection{Results for self-regulated disks}

Now, we  finally explore the chemical evolution of initially atomic disks in a self-regulated scenario with $Q=1$. In this regime, we focus again on the case with a central source, as otherwise the evolution would be very similar to the case of a generic disk model. We consider here central sources with masses of $1-10^3$~M$_\odot$, and characteristic accretion rates of $10^{-1}$~M$_\odot$~yr$^{-1}$. We  start with an initial surface density of $10$~g~cm$^{-2}$, which however adjusts early in the evolution due to the requirement of $Q=1$. Due to the generally higher surface and number densities in this scenario, we restrict the evolution here to a minimum radius of $1$~AU. In fact, considering stellar evolution models for supermassive stars, their typical extend may even exceed such a scale \citep{Hosokawa13}.  The overall models considered here are summarized in Table~\ref{tab:hotself}.

\begin{table}[htp]
\begin{center}
\begin{tabular}{c|c|c|c|c}
Model & $M_*$~[$M_\odot$] & $\dot{M}$~[M$_\odot$~yr$^{-1}$] & $\Sigma$~[g~cm$^{-2}$]  \\ \hline
HS0 & $1$ & $10^{-1}$ & $10$ \\
HS1 & $10$ & $10^{-1}$ & $10$ \\
HS2 & $100$ & $10^{-1}$ & $10$ \\
HS3 & $10^3$ & $10^{-1}$ & $10$ 
\end{tabular}
\end{center}
\caption{Models for a self-regulated disk with initially atomic gas.}
\label{tab:hotself}
\end{table}%

\begin{figure}[h]
\begin{center}
\includegraphics[scale=0.65]{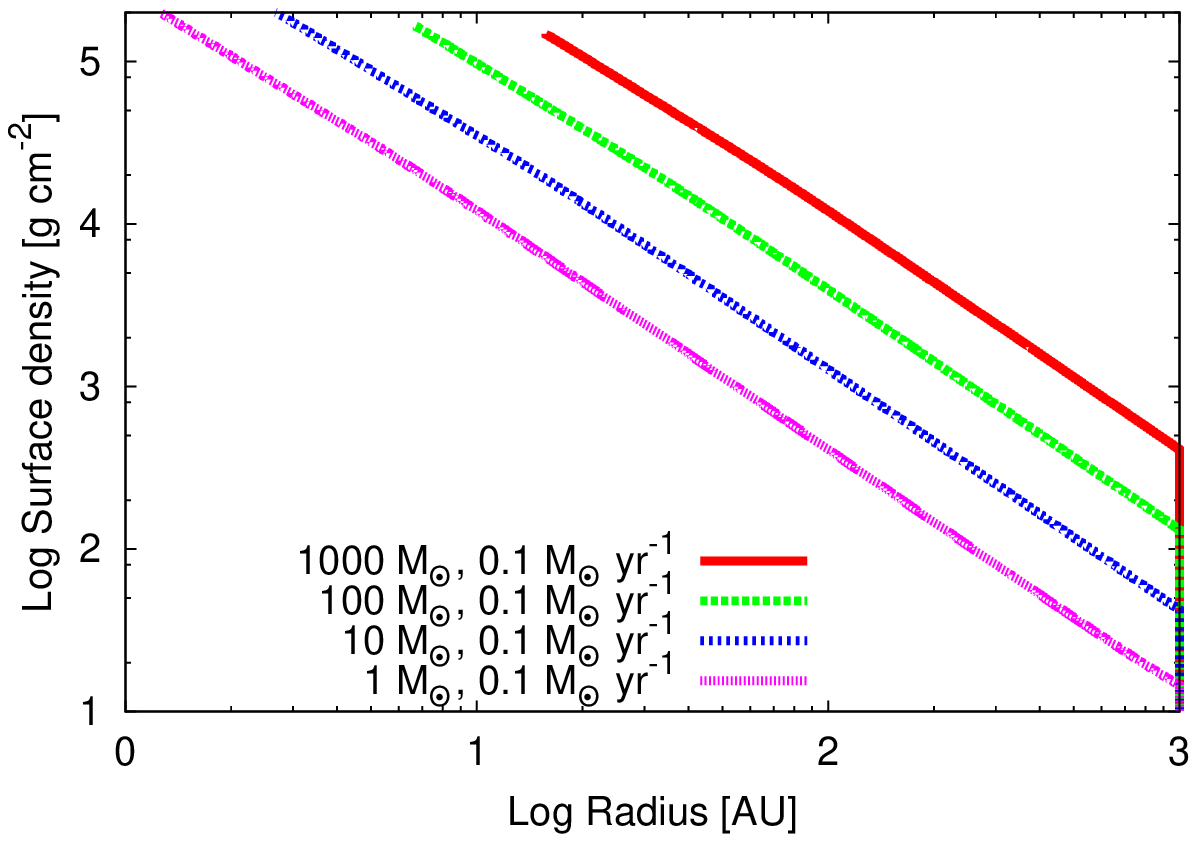}
\includegraphics[scale=0.65]{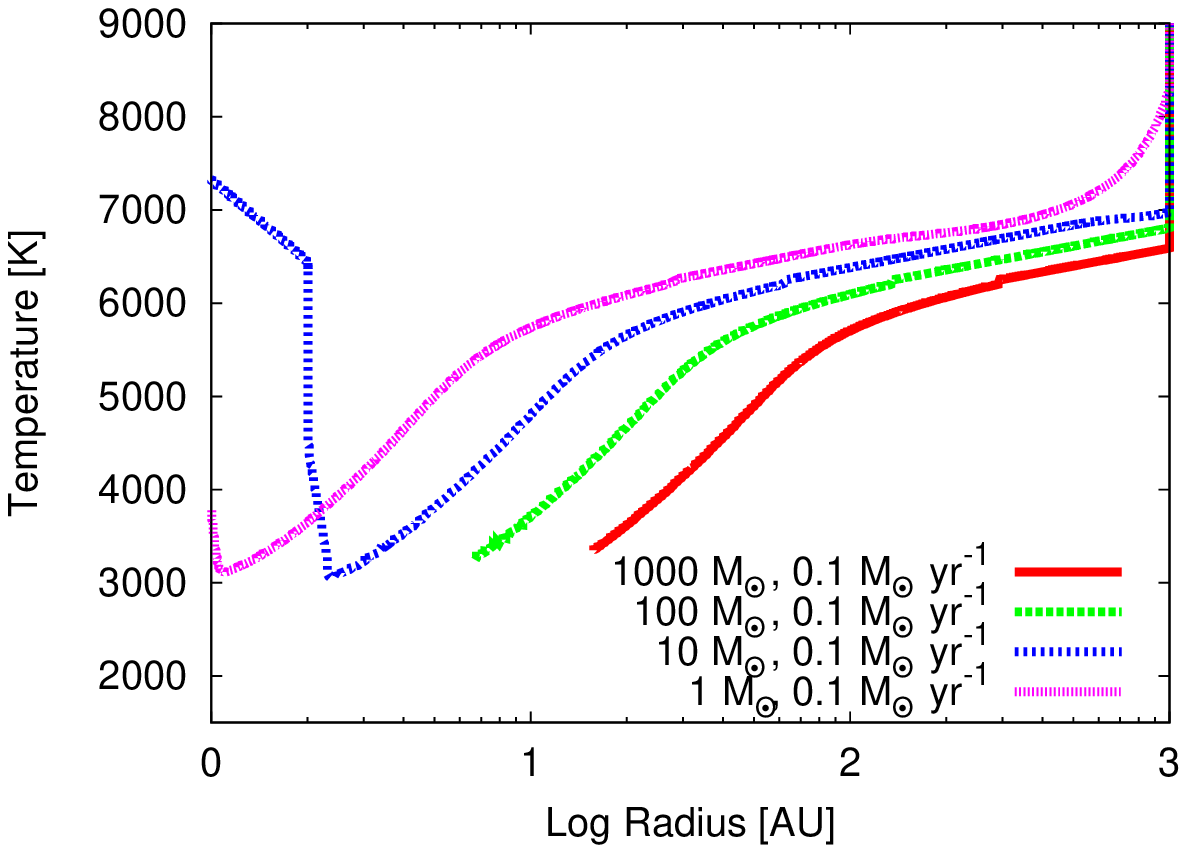}
\includegraphics[scale=0.65]{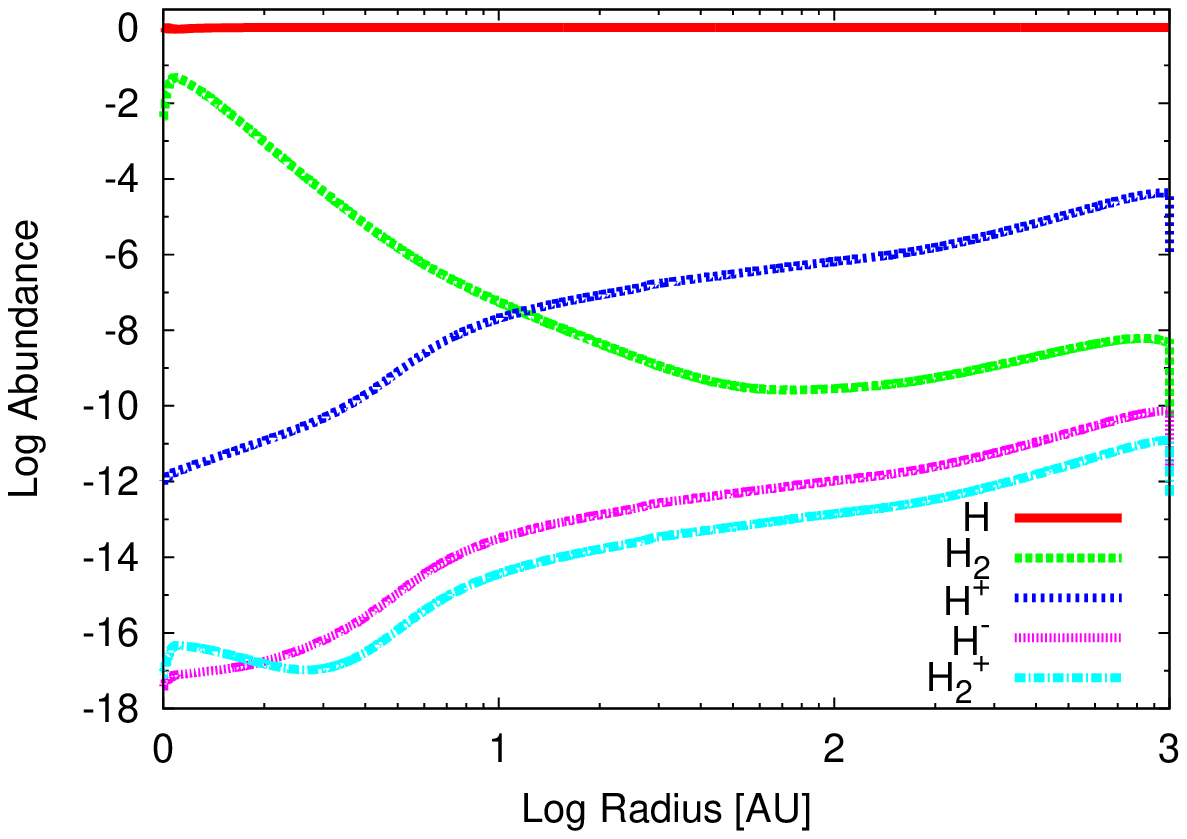}
\includegraphics[scale=0.65]{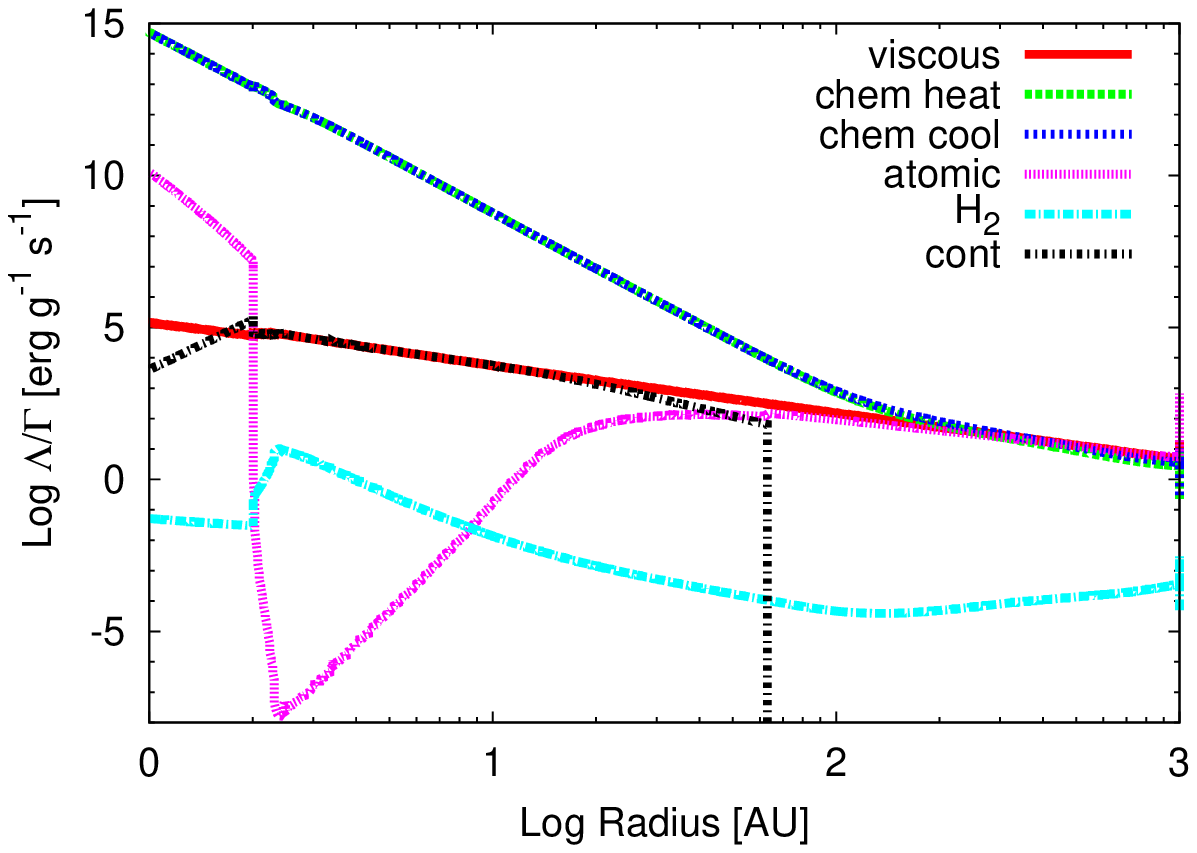}
\caption{Results for the self-regulated disk models with an initially atomic gas (Table~\ref{tab:hotself}). The species abundances and heating/cooling rates are given for the HS0 (self-regulated disk with central source of $1$~M$_\odot$ and an accretion rate of $10^{-1}$~M$_\odot$~yr$^{-1}$). We refer to the caption of Fig.~\ref{toomre} for the description of the panels. }
\label{hotq}
\end{center}
\end{figure}

The surface density of the gas as a function of radius is given in Fig.~\ref{hotq} (top panel), showing that the surface density profile  clearly steepens compared to a generic disk model, requiring a scaling as $\Sigma\propto R^{-1.5}$. Different from the case with initially molecular cooling, the profile strongly resembles a power-law with no particular features, as the temperature evolution is generally more gradual and mostly without strong transitions.  The temperature initially remains high, while some continuum cooling may set in between $60$ and $10$~AU (Fig.~\ref{hotq}, 2nd panel). The temperature then decreases more steeply towards $\sim3000$~K. For central sources with $1$ and $10$~M$_\odot$, we find that the temperature subsequently increases again, as a result of viscous heating and H$_2$ collisional dissociation. For higher central masses, the calculation becomes unstable at the point of this transition, due to the correspondingly higher heating and cooling rates. In principle, we expect however a similar evolution.

The chemical evolution is given for a characteristic case with a central source of $1$~M$_\odot$ and an accretion rate of $10^{-1}$~M$_\odot$~yr$^{-1}$, corresponding to model HS0 (Fig.~\ref{hotq}, 3rd panel). While the gas is atomic in the entire range, the H$_2$ abundance increases significantly from $\sim10^{-9}$ at $\sim50$~AU, reaching a peak value close to $0.1$ at $1$~AU. The ionization degree initially decreases only slightly with density between $1000$ and $10$~AU, and then drops more significantly towards $10^{-12}$  due to the lower temperatures. 

As shown in Fig.~\ref{hotq} (bottom panel), the viscous heating is balanced by atomic hydrogen cooling between $1000$ and $30$~AU. In the interior, the contribution of the continuum cooling is subsequently close to the viscous heating. In this regime, the chemical heating and cooling as well as the contributions of viscous heating and continuum cooling strongly balance each other, and the temperature decrease is driven by the molecular hydrogen cooling. The characteristic temperature for H$_2$ dissociation however depends on the density, and the critical point for H$_2$ dissociation is indeed reached at $R\sim1$~AU, leading to the dissociation of molecular hydrogen and a resulting increase of the temperature. 

Also in this case, the role of viscous heating is therefore  less pronounced for the initially atomic than for the initially molecular regime. However, due to the higher gas densities in the self-gravitating regime, the H$_2$ formation is enhanced for these cases, so that the cooling is initially somewhat enhanced, and the viscous heating prevents the gas from becoming molecular again in the central $1-10$~AU, therefore somewhat contributing to the stability in the very central region. This behavior concerns relatively small scales. A supermassive star with $\sim10^4$~M$_\odot$ may already have radii up to $\sim100$~AU \citep{Hosokawa13}, providing a natural cut-off for the inner disk. It thus appears likely that the viscous heating has only a minor effect in the initially atomic regime.

\section{Summary and discussion}\label{summary}

We have developed a one-zone framework to describe the chemical and thermal evolution of primordial disks depending on the mechanism which provides the viscosity of the disk. Assuming a sufficient viscosity is always available as a result of turbulence and/or magnetic fields \citep[see e.g.][]{Balbus99, Hawley00}, we have derived the generic disk model under the assumption of stationarity and axisymmetry {as well as a surface density profile $\Sigma\propto R^{-1}$, corresponding to a Mestel disk. In this framework,} the effective viscosity of the disk follows from the requirement to maintain the accretion rate and the transport of angular momentum, therefore providing the amount of viscous heating \citep[see e.g.][]{Lodato07, Inayoshi14, Latif15Disk}. We further distinguish between scenarios where the rotation profile is determined by the self-gravity of the disk, as well as scenarios where gravity is dominated by the central source. 

If rotation is dominated by the central source, it follows that the disk becomes gravitationally stable on smaller scales due to the relation $\Sigma\propto R^{-1}$. In particular in the regime of high accretion rates, one may however expect that self-gravity contributes signficantly to the transport of angular momentum \citep[e.g.][]{Begelman09}. We therefore consider also a self-regulated disk model, in which the disk viscosity is due to gravitational instabilities, therefore requiring a scenario with the Toomre Q$\sim1$. In these cases, we obtain a steeper profile for the surface density of the disk in a marginally stable state. We refer here to \citet{Lodato07} for a review on the properties of such disks. 

The resulting evolution has been explored for a range of stellar masses and accretion rates, considering both conventional Pop.~III star formation \citep[e.g.][]{Abel02, Bromm03, Yoshida08} as well as the formation of supermassive primordial stars \citep[e.g.][]{Lodato06, Regan09, Latif13d, Ferrara14}. As chemical initial conditions, we have both explored an initially molecular gas, as expected for instance in minihalos \citep[e.g.][]{Abel02, Bromm03, Yoshida08}, as well as an initially atomic gas. In more massive haloes, the chemical evolution depends on the ambient radiation background, and is typically molecular for moderate values and atomic for very strong values of the radiation background \citep[e.g.][]{Omukai01,Greif08, Shang09, Schleicher10b, Latif11b, Latif15X}. 

In case of the initially molecular gas, we find that viscous heating plays a crucial role in the presence of a central source and accretion rates of $\sim10^{-1}$~M$_\odot$~yr$^{-1}$, both for the generic and the self-regulated disk model. For a central star with about $10$~M$_\odot$, the viscous heating leads to the collisional dissociation of molecular hydrogen on scales of $\sim10$~AU, and the effect is further enhanced for supermassive stars with $\sim10^4$~M$_\odot$, where the molecular gas is dissociated already on scales of $\sim1000$~AU. The latter thus strongly contributes to the stabilization of the central region. Even if fragmentation occurs, the resulting clumps will be more massive, and are expected to rapidly migrate towards the central region \citep{Inayoshi14, Latif15}. {The transition towards an atomic gas also increases the opacity, which is typically of order $1$~cm$^2$~g$^{-1}$ at a temperature of $10^4$~K. The latter can have a relevant impact on the structure of the disk, and increase the overall stability.} However, fragmentation may be more efficient on larger scales where the gas is molecular, and also the migration time is increasing with the distance from the central object. It is thus certainly viable that fragmentation occurs on some scale, which could potentially contribute to the formation of a starburst ring around the central object \citep{Lodato06}. In such a case, the further accretion may be regulated by the interplay of gravitational instabilities with stellar feedback \citep{Kawakatu09, Wutschik13}.

We further explored the case where the gas is initially atomic. In case of the generic disk model, it then remains atomic, as collisional dissociation of molecular hydrogen is efficient at the temperatures of atomic hydrogen cooling \citep{Inayoshi12}. Due to the high temperatures, the overall cooling rates are considerably enhanced, and the effect of viscous heating is therefore less relevant in this regime, as also described by \citet{Ferrara13} for a hot metal-enriched gas. In the case of a self-gravitating disk model, we find that some H$_2$ may build up again towards the central region, due to the higher densities obtained in this regime. However, the characteristic temperatures are still above $\sim3000$~K throughout the evolution. As the characteristic temperature for H$_2$ collision dissociation decreases with increasing density, we  find that the H$_2$ dissociates again after an initial build-up. Overall, the gas essentially remains atomic for most of the evolution.

The potential impact of viscous heating is well-known for instance in the case of quasar accretion disks, where the resulting heat input leads to the presence of a hot gas over a large range of scales \citep[e.g.][]{Goodman03, Goodman04}. In the context of primordial star formation, such effects have not been strongly considered. The one-zone models exploring the chemical evolution have typically assumed a free-fall collapse \citep[e.g.][]{Omukai01, Omukai05, Glover08, Glover15}, while simulations were usually restricted to the early stages of disk formation. For instance, \citet{Clark11} were following the evolution until $\sim1$~M$_\odot$ of mass had turned into stars, while \citet{Greif11} stopped the evolution when a $\sim10$~M$_\odot$ star was formed, still with a moderate accretion rate of $10^{-2}$~M$_\odot$. In a follow-up investigation at higher resolution, they evolved the simulation until stellar masses of $\sim1$~M$_\odot$, and noted that heating both due to the accretion onto the protostar, as well as due to shocks close to the spiral arm has occured \citep{Greif12}. It is conceivable that these are the first signatures of viscous heating, even though still at an early stage. In the context of supermassive stars, we note that the majority of runs already started from an atomic gas, which typically remained atomic during the further evolution \citep[e.g.][]{Regan09, LatifBH, Latif13BHmass, Prieto13, Regan14, Becerra15}. The longer-term evolution in the presence of molecular cooling has been explored in simulations by \citet{Latif14Star} and \citet{LatifV15}, showing that high accretion rates can be maintained, but without resolving the scales considered here.

The results obtained here are particularly important for the formation of massive central object, as the enhanced temperature in the atomic gas increases the stability in the interior of the disk and helps to maintain a high accretion rate. The latter also alleviates the need of a strong external radiation field, typically expressed through a critical value $J_{\rm crit}$, as a means of providing an atomic gas. We expect these results to be particularly relevant in the case of massive central objects of at least $10$~M$_\odot$ and high accretion rates of $\sim10^{-1}$~M$_\odot$~yr$^{-1}$. The nature of the phenomenon discussed here, and in particular the characteristic scaling of the viscous heating rate as $R^{-3}$ leads us to the expectation that such a transition should generally occur, even though it may be shifted for instance if the surface density is enhanced and thus also the ability of the gas to cool. Even in the case of fragmentation and a more turbulent flow, we expect that gas velocities of the order of the Keplerian velocity will occur, leading to the formation of strong shocks and a transition towards the atomic cooling regime.

Of course, the models here are still based on simplifying assumptions, such as the stationarity and axisymmetry of the disk, which can only hold at an approximate level. It is therefore important to investigate the evolution in such disks in 3D simulations following the interplay of chemistry, heating and cooling along with the gravitational dynamics to assess the impact on fragmentation. For such time-dependent models, the viscosity of the disk can no longer be obtained from the assumption of stationarity, but one may have to adopt explicit parametrizations of the viscosity in self-gravitating disks as derived for instance by \citet{Rafikov15}. Due to the timescales involved in the evolution of such disks, we expect that the latter can not be pursued in cosmological simulations, but that independent studies are needed to investigate the stability of self-gravitating primordial disks  with already a massive protostar. Depending on the mass and accretion rate of the protostar, also radiation feedback may become relevant, even though it may predominantly affect the regions outside the disk \citep[e.g.][]{Hosokawa11, Hirano14}. In the case of rapid accretion, it was  found that the UV feedback from protostars is likely negligible, as they are expected to have cool atmospheres like a red giant \citep{Hosokawa13, Schleicher13b}. Previous investigations further have shown that $\alpha$ disk models with an optically thin cooling are thermally unstable \citep{Cannizzo84, Kato98}. {We note that the models pursued here are more complex, and in particular do not employ a constant $\alpha$, as the disk viscosity is derived from Eq.~\ref{nu}. It is nevertheless conceivable that thermal instabilities may play a relevant role in their evolution, which needs to be explored in further detail.}

Beyond the primordial case, the influence of dust cooling is potentially significant at high densities \citep[e.g.][]{Schneider03, Omukai05, Schneider06, Omukai08, Cazaux09, Dopcke11, Dopcke12, Schneider12}. However, most of these investigations have employed one-zone models assuming a free-fall collapse, while we have shown here that the heating mechanism is considerably different in a rotating disk around a central protostar. In this regime, the role of dust cooling requires further investigations, to understand  the fragmentation behavior at the later stages of the evolution. The latter can be important both for the formation of the first low-mass stars \citep[e.g.][]{Schneider12, Klessen12}, but also for the formation scenarios of massive black holes \citep{Omukai08, Ferrara13}. Even if fragmentation occurs, a black hole may possibly form, as long as the cluster is fed via substantial inflows and gravitational instabilities occur \citep{Alexander14, Latif15dust}.

\begin{acknowledgements}
DRGS and SB  thank for funding through the DFG priority program ''The Physics of the Interstellar Medium'' (projects SCHL 1964/1-1, SCHL 1964/1-2 and BO 4113/1-2). ML received funding from the ''BLACK'' project funded via the European Research Council under the European Community's Seventh Framework Programme (FP7/2007-2013 Grant Agreement no. 614199, project ``BLACK''). TG acknowledges the Centre for Star and Planet Formation funded by the Danish National Research Foundation.
\end{acknowledgements}


\begin{thebibliography}{97}
\expandafter\ifx\csname natexlab\endcsname\relax\def\natexlab#1{#1}\fi

\bibitem[{{Abel} {et~al.}(2002){Abel}, {Bryan}, \& {Norman}}]{Abel02}
{Abel}, T., {Bryan}, G.~L., \& {Norman}, M.~L. 2002, Science, 295, 93

\bibitem[{{Agarwal} {et~al.}(2014){Agarwal}, {Dalla Vecchia}, {Johnson},
  {Khochfar}, \& {Paardekooper}}]{Agarwal14}
{Agarwal}, B., {Dalla Vecchia}, C., {Johnson}, J.~L., {Khochfar}, S., \&
  {Paardekooper}, J.-P. 2014, \mnras, 443, 648

\bibitem[{{Agarwal} {et~al.}(2015){Agarwal}, {Smith}, {Glover}, {Natarajan}, \&
  {Khochfar}}]{Agarwal15}
{Agarwal}, B., {Smith}, B., {Glover}, S., {Natarajan}, P., \& {Khochfar}, S.
  2015, ArXiv e-prints

\bibitem[{{Alexander} \& {Natarajan}(2014)}]{Alexander14}
{Alexander}, T. \& {Natarajan}, P. 2014, Science, 345, 1330

\bibitem[{{Ba{\~n}ados} {et~al.}(2014){Ba{\~n}ados}, {Venemans}, {Morganson},
  {Decarli}, {Walter}, {Chambers}, {Rix}, {Farina}, {Fan}, {Jiang}, {McGreer},
  {De Rosa}, {Simcoe}, {Wei{\ss}}, {Price}, {Morgan}, {Burgett}, {Greiner},
  {Kaiser}, {Kudritzki}, {Magnier}, {Metcalfe}, {Stubbs}, {Sweeney}, {Tonry},
  {Wainscoat}, \& {Waters}}]{Banados14}
{Ba{\~n}ados}, E., {Venemans}, B.~P., {Morganson}, E., {et~al.} 2014, \aj, 148,
  14

\bibitem[{{Balbus} \& {Papaloizou}(1999)}]{Balbus99}
{Balbus}, S.~A. \& {Papaloizou}, J.~C.~B. 1999, \apj, 521, 650

\bibitem[{{Becerra} {et~al.}(2015){Becerra}, {Greif}, {Springel}, \&
  {Hernquist}}]{Becerra15}
{Becerra}, F., {Greif}, T.~H., {Springel}, V., \& {Hernquist}, L.~E. 2015,
  \mnras, 446, 2380

\bibitem[{{Beckwith} {et~al.}(1990){Beckwith}, {Sargent}, {Chini}, \&
  {Guesten}}]{Beckwith90}
{Beckwith}, S.~V.~W., {Sargent}, A.~I., {Chini}, R.~S., \& {Guesten}, R. 1990,
  \aj, 99, 924

\bibitem[{{Begelman} \& {Shlosman}(2009)}]{Begelman09}
{Begelman}, M.~C. \& {Shlosman}, I. 2009, \apjl, 702, L5

\bibitem[{{Bertin}(1997)}]{Bertin97}
{Bertin}, G. 1997, \apjl, 478, L71

\bibitem[{{Bovino} {et~al.}(2013){Bovino}, {Grassi}, {Latif}, \&
  {Schleicher}}]{Bovino13b}
{Bovino}, S., {Grassi}, T., {Latif}, M.~A., \& {Schleicher}, D.~R.~G. 2013,
  \mnras, 434, L36

\bibitem[{{Bovino} {et~al.}(2014{\natexlab{a}}){Bovino}, {Grassi},
  {Schleicher}, \& {Latif}}]{Bovino14}
{Bovino}, S., {Grassi}, T., {Schleicher}, D.~R.~G., \& {Latif}, M.~A.
  2014{\natexlab{a}}, \apjl, 790, L35

\bibitem[{{Bovino} {et~al.}(2014{\natexlab{b}}){Bovino}, {Latif}, {Grassi}, \&
  {Schleicher}}]{Bovino13c}
{Bovino}, S., {Latif}, M.~A., {Grassi}, T., \& {Schleicher}, D.~R.~G.
  2014{\natexlab{b}}, \mnras, 441, 2181

\bibitem[{{Bovino} {et~al.}(2014{\natexlab{c}}){Bovino}, {Schleicher}, \&
  {Grassi}}]{BovinoH2}
{Bovino}, S., {Schleicher}, D.~R.~G., \& {Grassi}, T. 2014{\natexlab{c}}, \aap,
  561, A13

\bibitem[{{Bromm} \& {Loeb}(2003)}]{Bromm03}
{Bromm}, V. \& {Loeb}, A. 2003, ApJ, 596, 34

\bibitem[{{Cannizzo} \& {Wheeler}(1984)}]{Cannizzo84}
{Cannizzo}, J.~K. \& {Wheeler}, J.~C. 1984, \apjs, 55, 367

\bibitem[{{Cazaux} \& {Spaans}(2009)}]{Cazaux09}
{Cazaux}, S. \& {Spaans}, M. 2009, \aap, 496, 365

\bibitem[{{Cen}(1992)}]{Cen92}
{Cen}, R. 1992, \apjs, 78, 341

\bibitem[{{Clark} {et~al.}(2008){Clark}, {Glover}, \& {Klessen}}]{Clark08}
{Clark}, P.~C., {Glover}, S.~C.~O., \& {Klessen}, R.~S. 2008, \apj, 672, 757

\bibitem[{{Clark} {et~al.}(2011){Clark}, {Glover}, {Smith}, {Greif}, {Klessen},
  \& {Bromm}}]{Clark11}
{Clark}, P.~C., {Glover}, S.~C.~O., {Smith}, R.~J., {et~al.} 2011, Science,
  331, 1040

\bibitem[{{Dijkstra} {et~al.}(2014){Dijkstra}, {Ferrara}, \&
  {Mesinger}}]{Dijkstra14}
{Dijkstra}, M., {Ferrara}, A., \& {Mesinger}, A. 2014, \mnras, 442, 2036

\bibitem[{{Dijkstra} {et~al.}(2008){Dijkstra}, {Haiman}, {Mesinger}, \&
  {Wyithe}}]{Dijkstra08}
{Dijkstra}, M., {Haiman}, Z., {Mesinger}, A., \& {Wyithe}, J.~S.~B. 2008,
  \mnras, 391, 1961

\bibitem[{{Dopcke} {et~al.}(2011){Dopcke}, {Glover}, {Clark}, \&
  {Klessen}}]{Dopcke11}
{Dopcke}, G., {Glover}, S.~C.~O., {Clark}, P.~C., \& {Klessen}, R.~S. 2011,
  ApJ, 729, L3

\bibitem[{{Dopcke} {et~al.}(2013){Dopcke}, {Glover}, {Clark}, \&
  {Klessen}}]{Dopcke12}
{Dopcke}, G., {Glover}, S.~C.~O., {Clark}, P.~C., \& {Klessen}, R.~S. 2013,
  \apj, 766, 103

\bibitem[{{Fan} {et~al.}(2004){Fan}, {Hennawi}, {Richards}, {Strauss},
  {Schneider}, {Donley}, {Young}, {Annis}, {Lin}, {Lampeitl}, {Lupton}, {Gunn},
  {Knapp}, {Brandt}, {Anderson}, {Bahcall}, {Brinkmann}, {Brunner}, {Fukugita},
  {Szalay}, {Szokoly}, \& {York}}]{Fan04}
{Fan}, X., {Hennawi}, J.~F., {Richards}, G.~T., {et~al.} 2004, \aj, 128, 515

\bibitem[{{Fan} {et~al.}(2006){Fan}, {Strauss}, {Richards}, {Hennawi},
  {Becker}, {White}, {Diamond-Stanic}, {Donley}, {Jiang}, {Kim}, {Vestergaard},
  {Young}, {Gunn}, {Lupton}, {Knapp}, {Schneider}, {Brandt}, {Bahcall},
  {Barentine}, {Brinkmann}, {Brewington}, {Fukugita}, {Harvanek}, {Kleinman},
  {Krzesinski}, {Long}, {Neilsen}, {Nitta}, {Snedden}, \& {Voges}}]{Fan06}
{Fan}, X., {Strauss}, M.~A., {Richards}, G.~T., {et~al.} 2006, \aj, 131, 1203

\bibitem[{{Fan~et~al.}(2001)}]{Fan01}
{Fan~et~al.} 2001, AJ, 122, 2833

\bibitem[{{Ferrara} {et~al.}(2013){Ferrara}, {Haardt}, \&
  {Salvaterra}}]{Ferrara13}
{Ferrara}, A., {Haardt}, F., \& {Salvaterra}, R. 2013, \mnras, 434, 2600

\bibitem[{{Ferrara} {et~al.}(2014){Ferrara}, {Salvadori}, {Yue}, \&
  {Schleicher}}]{Ferrara14}
{Ferrara}, A., {Salvadori}, S., {Yue}, B., \& {Schleicher}, D. 2014, \mnras,
  443, 2410

\bibitem[{{Forrey}(2013)}]{Forrey13}
{Forrey}, R.~C. 2013, \apjl, 773, L25

\bibitem[{{Fromang} {et~al.}(2004){Fromang}, {Balbus}, {Terquem}, \& {De
  Villiers}}]{Fromang04}
{Fromang}, S., {Balbus}, S.~A., {Terquem}, C., \& {De Villiers}, J.-P. 2004,
  \apj, 616, 364

\bibitem[{{Glover}(2015)}]{Glover15}
{Glover}, S.~C.~O. 2015, ArXiv e-prints 1504.00514

\bibitem[{{Glover} \& {Abel}(2008)}]{Glover08}
{Glover}, S.~C.~O. \& {Abel}, T. 2008, \mnras, 388, 1627

\bibitem[{{Goodman}(2003)}]{Goodman03}
{Goodman}, J. 2003, \mnras, 339, 937

\bibitem[{{Goodman} \& {Tan}(2004)}]{Goodman04}
{Goodman}, J. \& {Tan}, J.~C. 2004, \apj, 608, 108

\bibitem[{{Grassi} {et~al.}(2014){Grassi}, {Bovino}, {Schleicher}, {Prieto},
  {Seifried}, {Simoncini}, \& {Gianturco}}]{Grassi14}
{Grassi}, T., {Bovino}, S., {Schleicher}, D.~R.~G., {et~al.} 2014, \mnras, 439,
  2386

\bibitem[{{Greif}(2015)}]{Greif15}
{Greif}, T.~H. 2015, Computational Astrophysics and Cosmology, 2, 3

\bibitem[{{Greif} {et~al.}(2012){Greif}, {Bromm}, {Clark}, {Glover}, {Smith},
  {Klessen}, {Yoshida}, \& {Springel}}]{Greif12}
{Greif}, T.~H., {Bromm}, V., {Clark}, P.~C., {et~al.} 2012, \mnras, 424, 399

\bibitem[{{Greif} {et~al.}(2008){Greif}, {Johnson}, {Klessen}, \&
  {Bromm}}]{Greif08}
{Greif}, T.~H., {Johnson}, J.~L., {Klessen}, R.~S., \& {Bromm}, V. 2008,
  \mnras, 387, 1021

\bibitem[{{Greif} {et~al.}(2011){Greif}, {Springel}, {White}, {Glover},
  {Clark}, {Smith}, {Klessen}, \& {Bromm}}]{Greif11}
{Greif}, T.~H., {Springel}, V., {White}, S.~D.~M., {et~al.} 2011, ApJ, 737, 75

\bibitem[{{Hawley}(2000)}]{Hawley00}
{Hawley}, J.~F. 2000, \apj, 528, 462

\bibitem[{{Hirano} {et~al.}(2014){Hirano}, {Hosokawa}, {Yoshida}, {Umeda},
  {Omukai}, {Chiaki}, \& {Yorke}}]{Hirano14}
{Hirano}, S., {Hosokawa}, T., {Yoshida}, N., {et~al.} 2014, \apj, 781, 60

\bibitem[{{Hosokawa} {et~al.}(2011){Hosokawa}, {Omukai}, {Yoshida}, \&
  {Yorke}}]{Hosokawa11}
{Hosokawa}, T., {Omukai}, K., {Yoshida}, N., \& {Yorke}, H.~W. 2011, Science,
  334, 1250

\bibitem[{{Hosokawa} {et~al.}(2013){Hosokawa}, {Yorke}, {Inayoshi}, {Omukai},
  \& {Yoshida}}]{Hosokawa13}
{Hosokawa}, T., {Yorke}, H.~W., {Inayoshi}, K., {Omukai}, K., \& {Yoshida}, N.
  2013, \apj, 778, 178

\bibitem[{{Inayoshi} \& {Haiman}(2014)}]{Inayoshi14}
{Inayoshi}, K. \& {Haiman}, Z. 2014, \mnras, 445, 1549

\bibitem[{{Inayoshi} \& {Omukai}(2012)}]{Inayoshi12}
{Inayoshi}, K. \& {Omukai}, K. 2012, \mnras, 422, 2539

\bibitem[{{Kato} {et~al.}(1998){Kato}, {Fukue}, \& {Mineshige}}]{Kato98}
{Kato}, S., {Fukue}, J., \& {Mineshige}, S., eds. 1998, {Black-hole accretion
  disks}

\bibitem[{{Kawakatu} \& {Wada}(2009)}]{Kawakatu09}
{Kawakatu}, N. \& {Wada}, K. 2009, \apj, 706, 676

\bibitem[{{Klessen} {et~al.}(2012){Klessen}, {Glover}, \& {Clark}}]{Klessen12}
{Klessen}, R.~S., {Glover}, S.~C.~O., \& {Clark}, P.~C. 2012, \mnras, 421, 3217

\bibitem[{{Koushiappas} {et~al.}(2004){Koushiappas}, {Bullock}, \&
  {Dekel}}]{Koushiappas04}
{Koushiappas}, S.~M., {Bullock}, J.~S., \& {Dekel}, A. 2004, MNRAS, 354, 292

\bibitem[{{Latif} {et~al.}(2015{\natexlab{a}}){Latif}, {Bovino}, {Grassi},
  {Schleicher}, \& {Spaans}}]{Latif15X}
{Latif}, M.~A., {Bovino}, S., {Grassi}, T., {Schleicher}, D.~R.~G., \&
  {Spaans}, M. 2015{\natexlab{a}}, \mnras, 446, 3163

\bibitem[{{Latif} {et~al.}(2014{\natexlab{a}}){Latif}, {Bovino}, {Van Borm},
  {Grassi}, {Schleicher}, \& {Spaans}}]{Latif14UV}
{Latif}, M.~A., {Bovino}, S., {Van Borm}, C., {et~al.} 2014{\natexlab{a}},
  \mnras, 443, 1979

\bibitem[{{Latif} {et~al.}(2015{\natexlab{b}}){Latif}, {Omukai}, {Habouzit},
  {Schleicher}, \& {Volonteri}}]{Latif15dust}
{Latif}, M.~A., {Omukai}, K., {Habouzit}, M., {Schleicher}, D.~R.~G., \&
  {Volonteri}, M. 2015{\natexlab{b}}, MNRAS, submitted (ArXiv e-prints
  1509.07034)

\bibitem[{{Latif} \& {Schleicher}(2015{\natexlab{a}})}]{Latif15Disk}
{Latif}, M.~A. \& {Schleicher}, D.~R.~G. 2015{\natexlab{a}}, \mnras, 449, 77

\bibitem[{{Latif} \& {Schleicher}(2015{\natexlab{b}})}]{Latif15}
{Latif}, M.~A. \& {Schleicher}, D.~R.~G. 2015{\natexlab{b}}, A\&A, accepted
  (ArXiv e-prints 1411.5902)

\bibitem[{{Latif} {et~al.}(2014{\natexlab{b}}){Latif}, {Schleicher}, {Bovino},
  {Grassi}, \& {Spaans}}]{Latif14Star}
{Latif}, M.~A., {Schleicher}, D.~R.~G., {Bovino}, S., {Grassi}, T., \&
  {Spaans}, M. 2014{\natexlab{b}}, \apj, 792, 78

\bibitem[{{Latif} {et~al.}(2013{\natexlab{a}}){Latif}, {Schleicher}, {Schmidt},
  \& {Niemeyer}}]{LatifBH}
{Latif}, M.~A., {Schleicher}, D.~R.~G., {Schmidt}, W., \& {Niemeyer}, J.
  2013{\natexlab{a}}, \mnras, 433, 1607

\bibitem[{{Latif} {et~al.}(2013{\natexlab{b}}){Latif}, {Schleicher}, {Schmidt},
  \& {Niemeyer}}]{Latif13d}
{Latif}, M.~A., {Schleicher}, D.~R.~G., {Schmidt}, W., \& {Niemeyer}, J.
  2013{\natexlab{b}}, \mnras, 433, 1607

\bibitem[{{Latif} {et~al.}(2013{\natexlab{c}}){Latif}, {Schleicher}, {Schmidt},
  \& {Niemeyer}}]{Latif13PopIII}
{Latif}, M.~A., {Schleicher}, D.~R.~G., {Schmidt}, W., \& {Niemeyer}, J.
  2013{\natexlab{c}}, \apjl, 772, L3

\bibitem[{{Latif} {et~al.}(2013{\natexlab{d}}){Latif}, {Schleicher}, {Schmidt},
  \& {Niemeyer}}]{Latif13BHmass}
{Latif}, M.~A., {Schleicher}, D.~R.~G., {Schmidt}, W., \& {Niemeyer}, J.~C.
  2013{\natexlab{d}}, \mnras, 436, 2989

\bibitem[{{Latif} {et~al.}(2011){Latif}, {Schleicher}, {Spaans}, \&
  {Zaroubi}}]{Latif11b}
{Latif}, M.~A., {Schleicher}, D.~R.~G., {Spaans}, M., \& {Zaroubi}, S. 2011,
  A\&A, 532, A66

\bibitem[{{Latif} \& {Volonteri}(2015)}]{LatifV15}
{Latif}, M.~A. \& {Volonteri}, M. 2015, ArXiv e-prints 1504.00263

\bibitem[{{Lenzuni} {et~al.}(1991){Lenzuni}, {Chernoff}, \&
  {Salpeter}}]{Lenzuni91}
{Lenzuni}, P., {Chernoff}, D.~F., \& {Salpeter}, E.~E. 1991, \apjs, 76, 759

\bibitem[{{Lodato}(2007)}]{Lodato07}
{Lodato}, G. 2007, Nuovo Cimento Rivista Serie, 30, 293

\bibitem[{{Lodato} \& {Natarajan}(2006)}]{Lodato06}
{Lodato}, G. \& {Natarajan}, P. 2006, MNRAS, 371, 1813

\bibitem[{{Mayer} {et~al.}(2014){Mayer}, {Fiacconi}, {Bonoli}, {Quinn},
  {Roskar}, {Shen}, \& {Wadsley}}]{Mayer14}
{Mayer}, L., {Fiacconi}, D., {Bonoli}, S., {et~al.} 2014, ArXiv e-prints
  1411.5683

\bibitem[{{Mayer} {et~al.}(2010){Mayer}, {Kazantzidis}, {Escala}, \&
  {Callegari}}]{Mayer10}
{Mayer}, L., {Kazantzidis}, S., {Escala}, A., \& {Callegari}, S. 2010, \nat,
  466, 1082

\bibitem[{{Mortlock~et~al.}(2011)}]{Mortlock11}
{Mortlock~et~al.} 2011, Nature, 474, 616

\bibitem[{{Natarajan}(2011)}]{Natarajan11b}
{Natarajan}, P. 2011, {The formation and evolution of massive black hole seeds
  in the early Universe}, ed. D.~J. {Saikia} \& V.~{Trimble} (World Scientific
  Publishing Co), 191--206

\bibitem[{{Omukai}(2000)}]{Omukai00}
{Omukai}, K. 2000, \apj, 534, 809

\bibitem[{{Omukai}(2001)}]{Omukai01}
{Omukai}, K. 2001, ApJ, 546, 635

\bibitem[{{Omukai} {et~al.}(2008){Omukai}, {Schneider}, \& {Haiman}}]{Omukai08}
{Omukai}, K., {Schneider}, R., \& {Haiman}, Z. 2008, \apj, 686, 801

\bibitem[{{Omukai} {et~al.}(2005){Omukai}, {Tsuribe}, {Schneider}, \&
  {Ferrara}}]{Omukai05}
{Omukai}, K., {Tsuribe}, T., {Schneider}, R., \& {Ferrara}, A. 2005, \apj, 626,
  627

\bibitem[{{Peters} {et~al.}(2014){Peters}, {Schleicher}, {Smith}, {Schmidt}, \&
  {Klessen}}]{Peters14}
{Peters}, T., {Schleicher}, D.~R.~G., {Smith}, R.~J., {Schmidt}, W., \&
  {Klessen}, R.~S. 2014, \mnras, 442, 3112

\bibitem[{{Prieto} {et~al.}(2013){Prieto}, {Jimenez}, \& {Haiman}}]{Prieto13}
{Prieto}, J., {Jimenez}, R., \& {Haiman}, Z. 2013, \mnras, 436, 2301

\bibitem[{{Rafikov}(2015)}]{Rafikov15}
{Rafikov}, R.~R. 2015, ArXiv e-prints 1501.04980

\bibitem[{{Regan} \& {Haehnelt}(2009)}]{Regan09}
{Regan}, J.~A. \& {Haehnelt}, M.~G. 2009, MNRAS, 396, 343

\bibitem[{{Regan} {et~al.}(2014){Regan}, {Johansson}, \& {Haehnelt}}]{Regan14}
{Regan}, J.~A., {Johansson}, P.~H., \& {Haehnelt}, M.~G. 2014, \mnras, 439,
  1160

\bibitem[{{Safranek-Shrader} {et~al.}(2014){Safranek-Shrader},
  {Milosavljevi{\'c}}, \& {Bromm}}]{Shrader14}
{Safranek-Shrader}, C., {Milosavljevi{\'c}}, M., \& {Bromm}, V. 2014, \mnras,
  438, 1669

\bibitem[{{Safranek-Shrader} {et~al.}(2015){Safranek-Shrader}, {Montgomery},
  {Milosavljevic}, \& {Bromm}}]{Shrader15}
{Safranek-Shrader}, C., {Montgomery}, M., {Milosavljevic}, M., \& {Bromm}, V.
  2015, ArXiv e-prints 1501.03212

\bibitem[{{Schleicher} {et~al.}(2013){Schleicher}, {Palla}, {Ferrara}, {Galli},
  \& {Latif}}]{Schleicher13b}
{Schleicher}, D.~R.~G., {Palla}, F., {Ferrara}, A., {Galli}, D., \& {Latif}, M.
  2013, \aap, 558, A59

\bibitem[{{Schleicher} {et~al.}(2010){Schleicher}, {Spaans}, \&
  {Glover}}]{Schleicher10b}
{Schleicher}, D.~R.~G., {Spaans}, M., \& {Glover}, S.~C.~O. 2010, ApJL, 712,
  L69

\bibitem[{{Schneider} {et~al.}(2003){Schneider}, {Ferrara}, {Salvaterra},
  {Omukai}, \& {Bromm}}]{Schneider03}
{Schneider}, R., {Ferrara}, A., {Salvaterra}, R., {Omukai}, K., \& {Bromm}, V.
  2003, Nature, 422, 869

\bibitem[{{Schneider} {et~al.}(2012){Schneider}, {Omukai}, {Limongi},
  {Ferrara}, {Salvaterra}, {Chieffi}, \& {Bianchi}}]{Schneider12}
{Schneider}, R., {Omukai}, K., {Limongi}, M., {et~al.} 2012, \mnras, 423, L60

\bibitem[{{Schneider} {et~al.}(2006){Schneider}, {Salvaterra}, {Ferrara}, \&
  {Ciardi}}]{Schneider06}
{Schneider}, R., {Salvaterra}, R., {Ferrara}, A., \& {Ciardi}, B. 2006, \mnras,
  369, 825

\bibitem[{{Sethi} {et~al.}(2010){Sethi}, {Haiman}, \& {Pandey}}]{Sethi10}
{Sethi}, S., {Haiman}, Z., \& {Pandey}, K. 2010, \apj, 721, 615

\bibitem[{{Shang} {et~al.}(2010{\natexlab{a}}){Shang}, {Bryan}, \&
  {Haiman}}]{Shang10}
{Shang}, C., {Bryan}, G.~L., \& {Haiman}, Z. 2010{\natexlab{a}}, \mnras, 402,
  1249

\bibitem[{{Shang} {et~al.}(2010{\natexlab{b}}){Shang}, {Bryan}, \&
  {Haiman}}]{Shang09}
{Shang}, C., {Bryan}, G.~L., \& {Haiman}, Z. 2010{\natexlab{b}}, \mnras, 402,
  1249

\bibitem[{{Sugimura} {et~al.}(2014){Sugimura}, {Omukai}, \&
  {Inoue}}]{Sugimura14}
{Sugimura}, K., {Omukai}, K., \& {Inoue}, A.~K. 2014, \mnras, 445, 544

\bibitem[{{Susa} {et~al.}(2014){Susa}, {Hasegawa}, \& {Tominaga}}]{Susa14}
{Susa}, H., {Hasegawa}, K., \& {Tominaga}, N. 2014, \apj, 792, 32

\bibitem[{{Tanaka} \& {Omukai}(2014)}]{Tanaka14}
{Tanaka}, K.~E.~I. \& {Omukai}, K. 2014, \mnras, 439, 1884

\bibitem[{{Toomre}(1964)}]{Toomre64}
{Toomre}, A. 1964, \apj, 139, 1217

\bibitem[{{Turk} {et~al.}(2009){Turk}, {Abel}, \& {O'Shea}}]{Turk10}
{Turk}, M.~J., {Abel}, T., \& {O'Shea}, B. 2009, Science, 325, 601

\bibitem[{{Van Borm} \& {Spaans}(2013)}]{Borm13}
{Van Borm}, C. \& {Spaans}, M. 2013, \aap, 553, L9

\bibitem[{{Volonteri} \& {Bellovary}(2012)}]{Volonteri12}
{Volonteri}, M. \& {Bellovary}, J. 2012, Reports on Progress in Physics, 75,
  124901

\bibitem[{{Wutschik} {et~al.}(2013){Wutschik}, {Schleicher}, \&
  {Palmer}}]{Wutschik13}
{Wutschik}, S., {Schleicher}, D.~R.~G., \& {Palmer}, T.~S. 2013, \aap, 560, A34

\bibitem[{{Yoshida} {et~al.}(2008){Yoshida}, {Omukai}, \&
  {Hernquist}}]{Yoshida08}
{Yoshida}, N., {Omukai}, K., \& {Hernquist}, L. 2008, Science, 321, 669

\end{thebibliography}

\end{document}